\documentclass[journal]{IEEEtran}
\ifCLASSINFOpdf
\else
\fi
\usepackage{mathtools}
\usepackage{color}
\usepackage{algorithm}
\usepackage[noend]{algpseudocode}

\usepackage{times}
\usepackage{epsfig}
\usepackage{graphicx}
\usepackage{amssymb}
\usepackage{romannum}
\usepackage{color}

\usepackage{amsmath}
\usepackage{breqn}
\usepackage{multirow}
\usepackage{subfigure}

\begin{document}
\title{Semi-supervised Learning for COVID-19 Image Classification via ResNet}
\author{Lucy Nwosu,~Xiangfang Li,~Lijun Qian,~\IEEEmembership{Senior Member,~IEEE,}~Seungchan Kim,~\IEEEmembership{Member,~IEEE},~Xishuang Dong,~\IEEEmembership{Member,~IEEE}
\thanks{L. Nwosu is with the Center of Computational Systems Biology, Department of Electrical and Computer Engineering, Prairie View A\&M University, Texas A\&M University System, Prairie View, TX 77446, USA (e-mail: lnwosu@student.pvamu.edu).}
\thanks{X. Li is with the Center of Excellence in Research and Education for Big Military Data Intelligence (CREDIT Center) and the Center of Computational Systems Biology, Department of Electrical and Computer Engineering, Prairie View A\&M University, Texas A\&M University System, Prairie View, TX 77446, USA (e-mail: xili@pvamu.edu).}
\thanks{L. Qian is with the Center of Excellence in Research and Education for Big Military Data Intelligence (CREDIT Center) and the Center of Computational Systems Biology, Department of Electrical and Computer Engineering, Prairie View A\&M University, Texas A\&M University System, Prairie View, TX 77446, USA (e-mail: liqian@pvamu.edu).}
\thanks{S. Kim is with the Center of Computational Systems Biology, Department of Electrical and Computer Engineering, Prairie View A\&M University, Texas A\&M University System, Prairie View, TX 77446, USA (e-mail: sekim@pvamu.edu).}
\thanks{X. Dong is with the Center of Computational Systems Biology and the Center of Excellence in Research and Education for Big Military Data Intelligence (CREDIT Center), Department of Electrical and Computer Engineering, Prairie View A\&M University, Texas A\&M University System, Prairie View, TX 77446, USA (e-mail: xidong@pvamu.edu).}
}

\maketitle

\begin{abstract}
Coronavirus disease 2019 (COVID-19) is an ongoing global pandemic in over 200 countries and territories, which has resulted in a great public health concern across the international community. Analysis of X-ray imaging data can play a critical role in timely and accurate screening and fighting against COVID-19. Supervised deep learning has been successfully applied to recognize COVID-19 pathology from X-ray imaging datasets. However, it requires a substantial amount of annotated X-ray images to train models, which is often not applicable to data analysis for emerging events such as COVID-19 outbreak, especially in the early stage of the outbreak. To address this challenge, this paper proposes a two-path semi-supervised deep learning model, \textit{ssResNet}, based on Residual Neural Network (ResNet) for COVID-19 image classification, where two paths refer to a supervised path and an unsupervised path, respectively. Moreover, we design a weighted supervised loss that assigns higher weight for the minority classes in the training process to resolve the data imbalance. Experimental results on a large-scale of X-ray image dataset \textit{COVIDx} demonstrate that the proposed model can achieve promising performance even when trained on very few labeled training images.
\end{abstract}

\begin{IEEEkeywords}
COVID-19 Image Classification,~Semi-supervised Learning,~Residual Neural Network,~Joint Optimization,~Data Imbalance
\end{IEEEkeywords}

\section{Introduction}
\label{sec1}

%\section{Introduction}
\begin{figure*} []
	\begin{center}
		\includegraphics[width=0.75\linewidth]{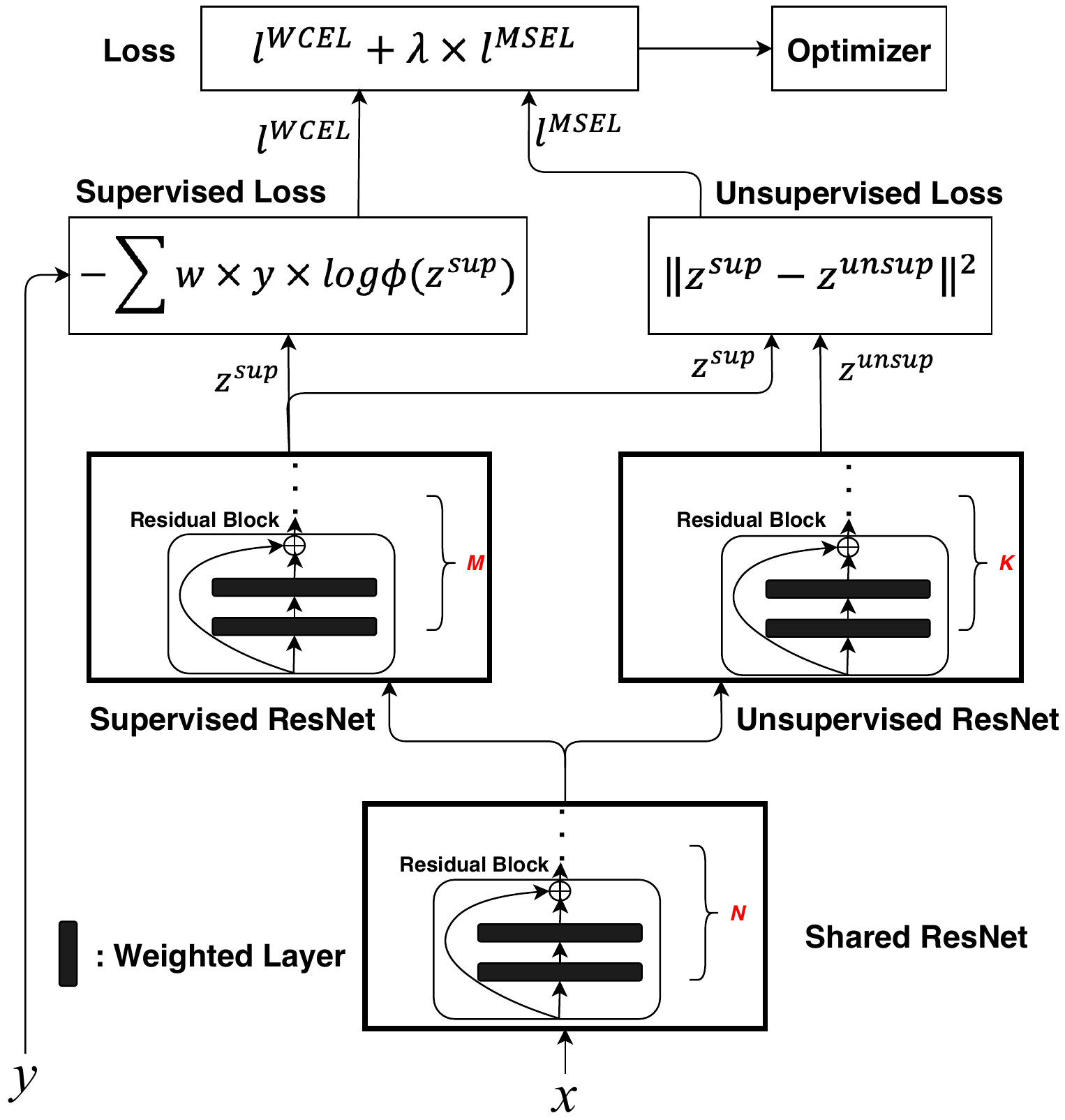}
		\caption{Framework of the proposed semi-supervised learning. Input $x$ is the medical image. Labels such as $y$ are available only for the labeled inputs. Shared ResNet will evaluate the input to obtain the low-level representations as inputs to supervised ResNet and unsupervised ResNet, where these three ResNets are built with residual blocks and  $N$, $M$, and $K$ are numbers of residual blocks for these three ResNets. Then $z^{sup}$ and $z^{unsup}$ are outputs from the supervised ResNet and the unsupervised ResNet, respectively. Moreover, $z^{sup}$ and $y$ will be applied to calculate a weighted cross entropy loss $l^{WCEL}$ whereas $z^{sup}$ and $z^{unsup}$ are used to calculate a mean squared error loss $l^{MSEL}$, where $w$ the weight to different classes of samples. We jointly optimize the combined losses, where $\lambda$ is the weight for unsupervised loss. $\oplus$ is the short-cut connection in the residue operation.}
		\label{Fig1:semisupervisedresnet}
	\end{center}
\end{figure*}

Coronavirus disease 2019 (COVID-19) outbreak has led to the heavy losses of the world's economy and life. To reduce the spread of COVID-19 and the death rate, it is essential to detect the disease at the early stage with effective and timely screening/testing and place COVID-19 infected patients in quarantine immediately~\cite{NYtimes,CDCguide}. Artificial intelligence (AI), an emerging technology for medical imaging processing, has actively contributed to the fight against COVID-19~\cite{bullock2020mapping}. Compared to the traditional imaging workflow that heavily relies on human interpretation, AI enables more safe, accurate and efficient imaging solutions. 

Recent AI-empowered applications in COVID-19 detection include the dedicated imaging platform, the lung and infection region segmentation, as well as the clinical assessment and diagnosis~\cite{shi2020review, gozes2020rapid, zheng2020deep}. Moreover, commercial products integrate AI to combat COVID-19 and demonstrate the capability of the AI technology~\cite{shi2020review}. All of these examples show the tremendous enthusiasm cast by the public for AI-empowered progress in the medical imaging field, especially during the ongoing COVID-19 pandemic. 

Regarding the COVD-19 research based on AI, COVID-19 image classification  becomes more and more attractive, which is to separate COVID-19 patients from non-COVID-19 subjects using the features extracted from medical images. Specially, supervised deep learning such as convolutional neural networks (CNN) has been very popular in this research area. For example, Wang \textit{et al.} proposed a 2D CNN supervised model to analyze delineated region patches to accomplish classification between COVID-19 and typical viral pneumonia~\cite{wang2020deep}. Similarly, Xu \textit{et al.} utilized candidate infection regions to complete COVID-19 classification via supervised ResNet-18~\cite{butt2020deep}. 

In addition, as a powerful deep learning model for medical image analysis, UNet~\cite{ronneberger2015u} was employed for COVID-19 image classification and segmentation. For example, Zheng \textit{et al.} employed UNet to obtain lung segmentation and predicted the probability of COVID-19 with 3D CNN on segmentation features~\cite{zheng2020deep}. Jin \textit{et al.}  proposed a UNet++ based segmentation model for locating lesions and built a ResNet-50 based classification model for COVID-19 diagnosis~\cite{jin2020ai}. Chen \textit{et al.} implemented COVID-19 classification with the patterns of segmented lesions extracted by supervised UNet++~\cite{zhou2018unet++, chen2020deep}. Moreover, they employed a 2D Deeplab model for the lung segmentation and a 2D ResNet-152 model for lung-mask slice based identification of positive COVID-19 cases~\cite{jin2020development}. Although supervised deep learning presents impressive performance on COVID-19 image classification, it requires a large amount of annotated medical images to train models, which is not practical with respect to limited data resources related to COVID-19, due to huge costs of labeling medical images, and labeling noise~\cite{litjens2017survey}.

To reduce the efforts on labeling medical images for COVID-19 image classification, we build a two-path semi-supervised deep learning model that is able to learn on both labeled and unlabeled medical images, based on residual neural networks (ResNet)~\cite{he2016deep}. ResNet is an artificial neural network developed by mimicking pyramidal cells in the cerebral cortex. It is to introduce a so-called ``identity shortcut connection" that skips one or more layers since stacking layers should not degrade the network performance. With ResNet, we implement a two-path semi-supervised learning model that is composed of three components, namely, shared ResNet, supervised ResNet, and unsupervised ResNet.  

Framework of the proposed model is shown in Fig.~\ref{Fig1:semisupervisedresnet}. One path is composed of a shared ResNet and a supervised ResNet while the other path consists of the shared ResNet and an unsupervised ResNet. All data (labeled and unlabeled data) will be evaluated to calculate the unsupervised loss that is the mean squared error loss (MSEL), while only labeled data will be used to calculate the supervised loss that is the cross entropy loss (CEL). Specifically, we design a weighted cross entropy loss (WCEL) that assigns more weight to the COVID-19 class for addressing the data imbalance. Reducing MSEL is to enhance the image representation while decreasing WCEL is to enhance classification performance. We validate the proposed model on a large-scale of X-ray image dataset \textit{COVIDx} and experimental results demonstrate the proposed model can accomplish COVID-19 image classification with promising performance even when trained on the extremely limited amount of labeled X-ray images.

The contributions in this study are below.

\begin{itemize}
\item We propose a semi-supervised deep learning model with ResNet through jointly training a supervised ResNet and an unsupervised ResNet. We observed  that the proposed model can learn on both unlabeled images and labeled images jointly for COVID-19 image classification with high performance.

\item The proposed model is validated on a large-scale COVID-19 image dataset. Experimental results indicate the proposed model is able to effectively recognize COVID-19 images by learning on very few labeled medical images, for example, less than 10\% samples in the training data, which meets the requirement of  few available labeled data from the medical domain for real applications~\cite{litjens2017survey}, especially for the cases at the early stage of such global pandemic.
\end{itemize}

\section{Proposed Methodology}
\label{sec2}

We propose a semi-supervised ResNet to address the challenge of lacking of labeled data for COVID-19  image classification, where the detailed framework is shown in Fig.~\ref{Fig1:semisupervisedresnet}. The shared ResNet will generate a new representation of  input $x$, where the new representation $z$ is given by

\begin{equation}
	z =  f_{pooling}(f_{Resblock_{N}}\cdots f_{Resblock_{1}}(x')) \; .
\label{Equ_output_shared_resnet}
\end{equation}
where 
\begin{equation}
	x' = f_{conv}(x) \; .
\label{Equ_output_conv}
\end{equation}

\begin{equation}
	f_{Resblock}(x') = x + f_{conv}(f_{conv}(x')) \; .
\label{Equ_output_shared_resblock}
\end{equation}

$f_{cov}(\cdot)$ is the convolutional operation.  $f_{Resblock}(\cdot)$ is the residual operation~\cite{he2016deep} and $f_{Resblock_{N}}\cdots f_{Resblock_{1}}(\cdot)$ refers to $N$ sequencing residual operations. $f_{pooling}(\cdot)$ is the pooling operation. Introducing this shared ResNet to the proposed model is inspired by deep multi-task learning~\cite{zhang2014facial, ruder2017overview},  since different tasks share a low-level feature representation extracted from the input $x$. In addition, the reason for learning low-level feature representations instead of directly using $x$ is that the original representation may not have enough expressive power for multiple tasks~\cite{bengio2013representation}. With the training data in all tasks, a more powerful representation can be learned for all tasks and this representation will improve performance. As shown in Fig.~\ref{Fig1:semisupervisedresnet}, we have two ``tasks" in our proposed model, namely, a supervised task and an unsupervised task, which is similar to the framework of deep multi-task learning. Therefore, the shared ResNet is necessary to feed the low-level representations to these two tasks.

The output $z$ from the shared ResNet is evaluated by two ResNets, namely, a supervised ResNet and an unsupervised ResNet. For the supervised ResNet, it is to learn the deep features of labeled samples. The output $z^{sup}$  of the supervised ResNet is given by 
\begin{equation}
	z^{sup} =  f_{pooling}^{sup}(f_{Resblock_{M}}^{sup}\cdots f_{Resblock_{1}}^{sup}(z')) \; .
\label{Equ_output_sup_resnet}
\end{equation}
where 
\begin{equation}
	z' = f_{conv}^{sup}(z) \; .
\label{Equ_output_sup_conv}
\end{equation}

\begin{equation}
	f_{Resblock}^{sup}(z') = z' + f_{conv}^{sup}(f_{conv}^{sup}(z')) \; .
\label{Equ_output_sup_resblock}
\end{equation}

We employ the same operations including the pooling operation $f_{pooling}^{sup}(\cdot)$, the convolutional operation $f_{conv}^{sup}(\cdot)$, and $M$ sequencing residual operations $f_{Resblock}^{sup}(\cdot)$. Moreover, we build the unsupervised ResNet to generate another representation of all inputs including labeled data and unlabeled data. This representation $z^{unsup}$ is given by

\begin{equation}
	z^{unsup} =  f_{pooling}^{unsup}(f_{Resblock_{K}}^{unsup}\cdots f_{Resblock_{1}}^{unsup}(z'')) \; .
\label{Equ_output_unsup_resnet}
\end{equation}
where 
\begin{equation}
	z'' = f_{conv}^{unsup}(z) \; .
\label{Equ_output_unsup_conv}
\end{equation}

\begin{equation}
	f_{Resblock}^{unsup}(z'') = z'' + f_{conv}^{unsup}(f_{conv}^{unsup}(z'')) \; .
\label{Equ_output_unsup_resblock}
\end{equation}

Similarly, we employ the pooling operation $f_{pooling}^{unsup}(\cdot)$, the convolutional operation $f_{conv}^{unsup}(\cdot)$, and  $K$ sequencing residual operation $f_{Resblock}^{unsup}(\cdot)$ to build the unsupervised ResNet. Then, we utilize those two vectors $z^{sup}$ and $z^{unsup}$ to calculate the weighted cross entropy loss (WCEL) and mean squared error loss (MSEL) for supervised and unsupervised paths, respectively. They are given by 
\begin{equation}
	l^{WCEL} =  -\sum w \times y \times log \phi{(z^{sup})} \; .
\label{Equ_loss1}
\end{equation}
\begin{equation}
	l^{MSEL}  = ||z^{sup} - z^{unsup}||^{2}\; .
\label{Equ_loss2}
\end{equation}
where $y$ is the ground truth of the input and $w$ is corresponding weight. $\phi(\cdot)$ is the softmax activation function.  $l^{WCEL}$ is the weighted cross entropy loss to account for the loss of labeled inputs. To enhance classification performance for the minority class (COVID-19 class), we assign more weight to COVID-19 class, where during the learning procedure the classifier will pay more attentions to the COVID-19 class so as to reduce the learning bias that is caused by data imbalance. 

$l^{MSEL}$ is to measure the differences between $z^{sup}$ and $z^{unsup}$. Since training ResNets with dropout regularization and gradient-based optimization is a stochastic process, the two ResNets will have different link weights after training. In other words, there will be differences between the two prediction vectors $z^{sup}$ and $z^{unsup}$ that are from these two ResNets (the supervised ResNet and the unsupervised ResNet). These differences can be treated as an error in the classification and thus minimizing this loss is a goal in the proposed model, which is inspired by $\Pi$ model~\cite{laine2016temporal}. 

Based on these two losses, the total loss is defined by 

\begin{equation}
	Loss = l^{WCEL} + \lambda \times l^{MSEL}\; .
\label{Equ_loss}
\end{equation}

where $\lambda$ is the weight for $ l^{MSEL}$. Training the proposed model is to optimize $Loss$ on the training data. At the beginning of training, the total loss and the learning gradients are dominated by the supervised loss component, i.e., the labeled data only. At later stage of training, unlabeled data will contribute more than labeled data. These processes are controlled by fine-tuning $\lambda$~\cite{laine2016temporal}. The detailed steps for learning of the proposed model is shown in  Algorithm \ref{Arg1_learning}. $f_{\theta_{shared}}(\cdot)$ is to learn the low-level features from the medical images. Parameters of the shared ResNet $\theta_{shared}$  include weights learned for the operations, namely, pooling operation $f_{pooling}(\cdot)$, convolutional operation $f_{conv}(\cdot)$, and residual operation $f_{Resblock}(\cdot)$. 

\begin{algorithm}[h!]
	\caption{Learning of Semi-supervised ResNet (SSResNet)}
	\begin{algorithmic}[1]
		\Require{training sample  $x_{i}$, the set of training samples  $S$, label $y_{i}$ for $x_{i}$ ($i \in S$) } 
		\For{$t$ in~[1, num epochs] }
 			 \For{each minibatch $B$}
			 	\State{$z_{i \in B} \gets f_{\theta_{shared}}({{x_{i \in B}}})$} $\triangleright$ shared  representation
      				\State{$z^{sup}_{i \in B} \gets f_{\theta_{sup}}({{z_{i \in B}}})$} $\triangleright$ supervised  representation
				\State{$z^{unsup}_{i \in B} \gets f_{\theta_{unsup}}({{z_{i \in B}}})$} $\triangleright$ unsupervised  representation
				\State{$l_{i \in B}^{WCEL} \gets -\frac{1}{|B|} \sum_{i \in B \cap S}{log \phi{(z^{sup}_{i})}[y_{i}w_{i}]}$}  $\triangleright$ supervised loss component
				\State{$l_{i \in B}^{MSEL} \gets \frac{1}{C|B|} \sum_{i \in B}{||z^{sup}_{i} - z^{unsup}_{i}||^{2}}$} $\triangleright$ unsupervised loss component
				\State{$Loss \gets l_{i \in B}^{WCEL} +  \lambda \times l_{i \in B}^{MSEL}$} $\triangleright$ total loss 
				\State{update $\theta_{shared}$,  $\theta_{sup}$, $\theta_{unsup}$ using optimizer , e.g., ADAM} 
 			 \EndFor
		\EndFor
	\Return{$\theta_{shared}$,  $\theta_{sup}$, $\theta_{unsup}$}
	\end{algorithmic}
	\label{Arg1_learning}
\end{algorithm}

After extracting low-level feature representations from inputs, we use $f_{\theta_{sup}}(\cdot)$ and $f_{\theta_{unsup}}(\cdot)$ to obtain higher level representations $z^{sup}$ and $z^{unsup}$, where $z^{sup}$ is used to complete COVID-19 classification. In addition,  $z^{sup}$ and $z^{unsup}$ are employed to enhance the image representations. Parameters of the supervised ResNet $\theta_{sup}$  include weights learned for the operations, namely, pooling operation $f_{pooling}^{sup}(\cdot)$, convolutional operation $f_{conv}^{sup}(\cdot)$, and residual operation $f_{Resblock}^{sup}(\cdot)$ while those of the unsupervised ResNet $\theta_{unsup}$  consist of weights learned for the operations, namely, pooling operation $f_{pooling}^{unsup}(\cdot)$, convolutional operation $f_{conv}^{unsup}(\cdot)$, and residual operation $f_{Resblock}^{unsup}(\cdot)$. Specially, in the training procedure, we overcome the data imbalance by assigning more weight $w_{i}$ to the minority class (COVID-19 class) of samples. Finally, we employ ADAM optimizer to jointly optimize the total loss.

%The proposed model combines the advantages of deep multi-task learning~\cite{zhang2014facial} and $\Pi$ model \cite{laine2016temporal}. However,  there exist significant differences. Compared to deep multi-task learning, the subtasks in the proposed model have two categories of learning, namely, supervised learning and unsupervised learning while there is only supervised learning in the deep multi-task learning. On the other hand, instead of using one path neural networks, we apply two independent RNNs to generate supervised and unsupervised outputs. Furthermore, the proposed model is more flexible as the two independent RNNs can be tuned in terms of specific goals. 

\section{Experiment}
\label{sec3}
\begin{figure*}[t] 
	\centering
	%\captionsetup{font={footnotesize }}
	\subfigure[normal]{
		\label{Fig-01-1}
		\includegraphics[height=1.85in]{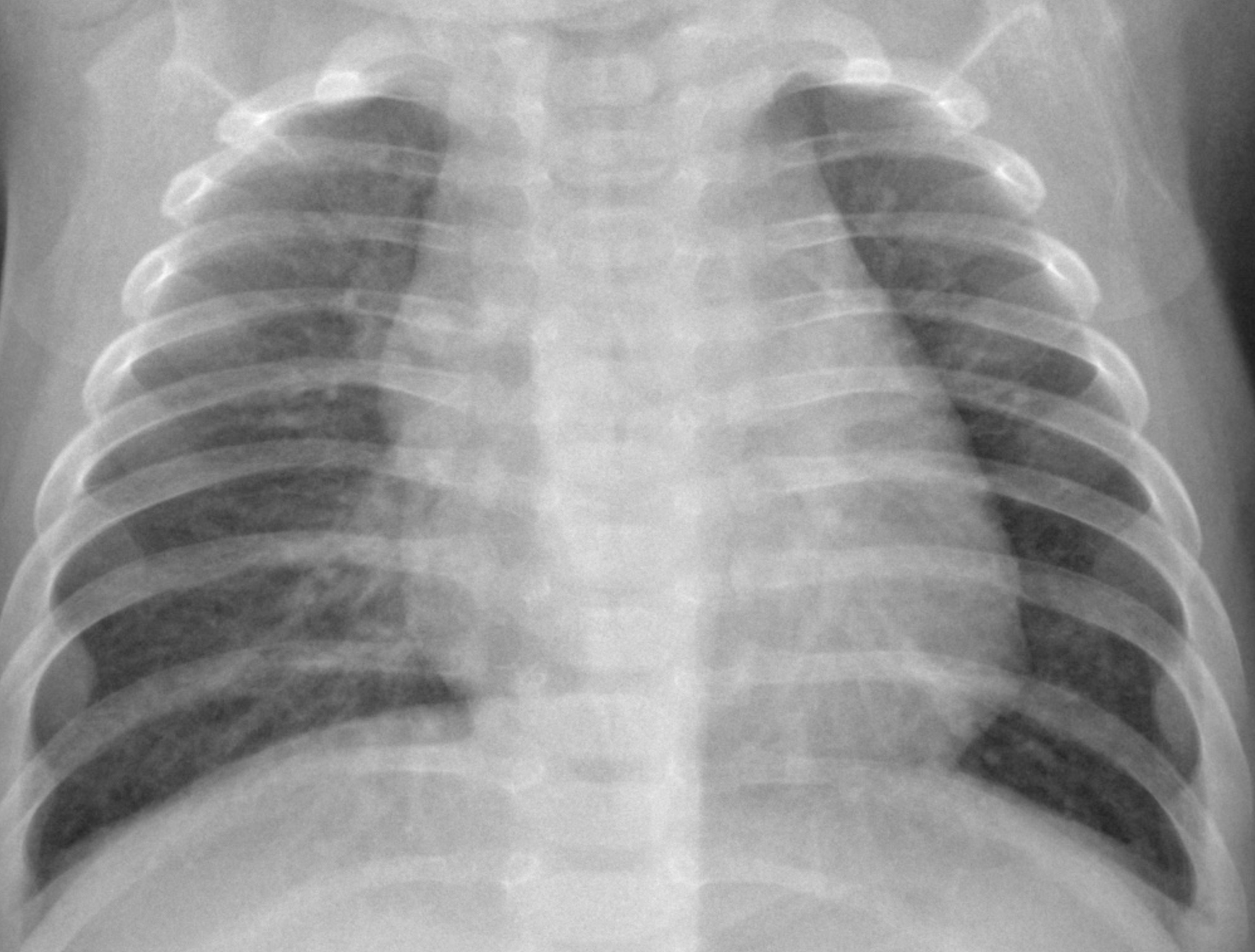}} 
	\hspace{0.02in}
	\subfigure[pneumonia]{
		\label{Fig-01-2}
		\includegraphics[height=1.85in]{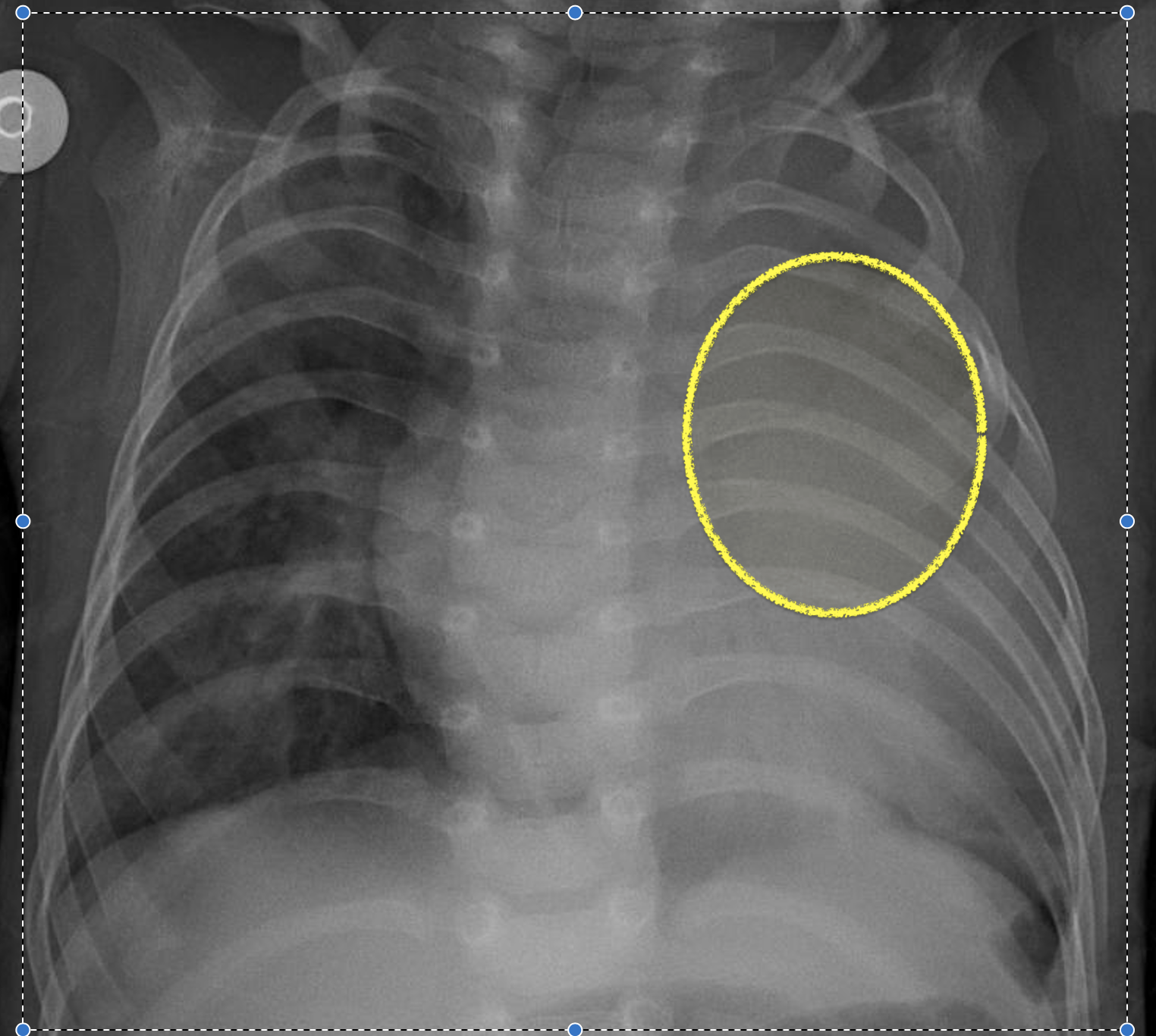}}
	\hspace{0.02in}
	\subfigure[COVID-19]{
		\label{Fig-01-3}
		\includegraphics[height=1.85in]{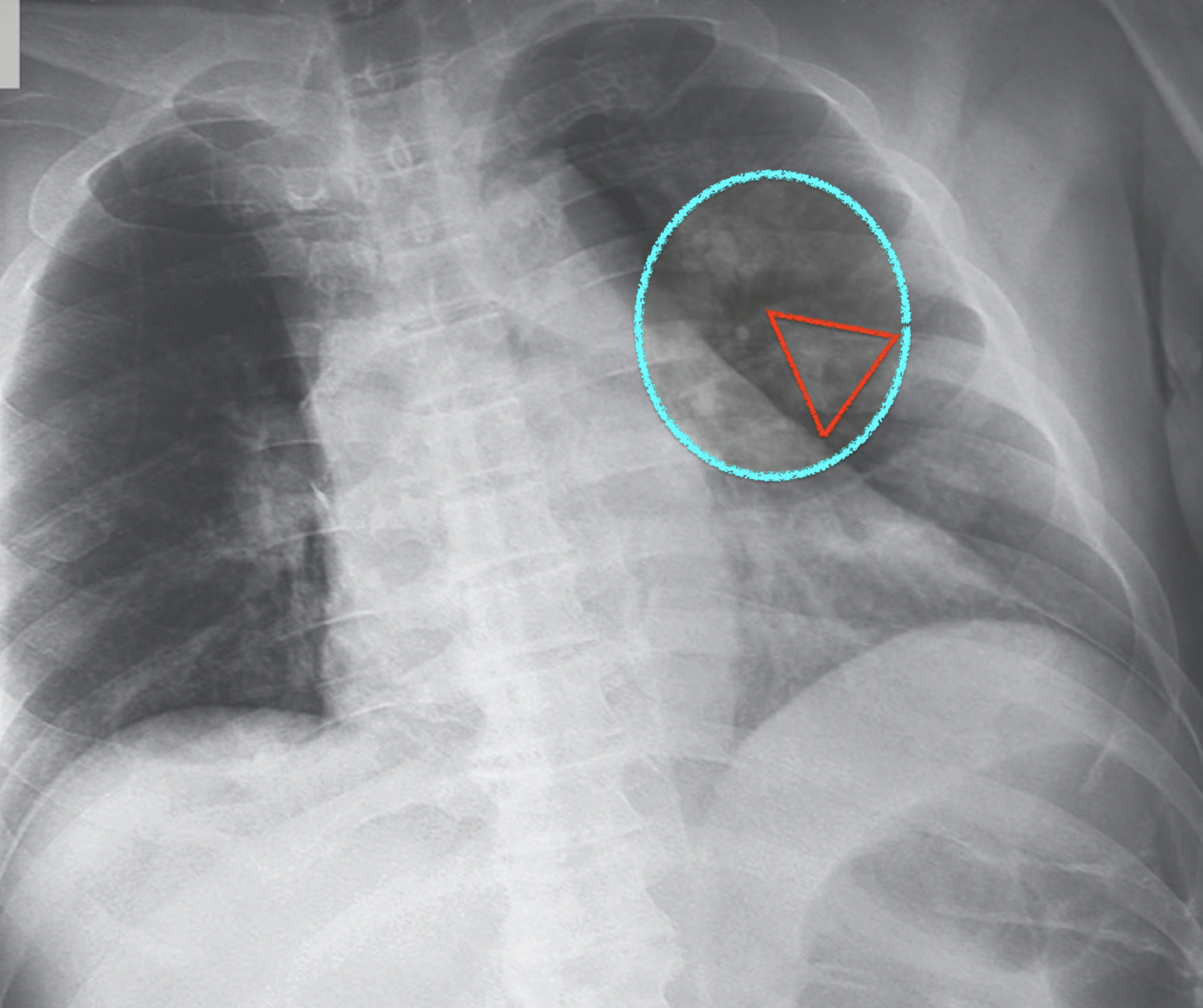}}
	\caption{Examples of chest radiography images belonging to normal, pneumonia, and COVID-19 classes are shown in (a), (b) and (c), respectively. Yellow circle locates infected regions of pneumonia for subfigure (b) while in subfigure (c) the red rectangle shape of region in the blue circle shows the potential infected areas of COVID-19.} 
	\label{ExampleImagesCOVIDX}
\end{figure*} 

\subsection{Dataset}

%Currently, COVID-19 pandemic is still continuing to affect the health and well-being of the global population. A critical step to fight against COVID-19 is to test and locate infected patients with effective and efficient tools. One of potential methods is to examine radiological images using chest radiography. From results in the previous studies, patients presenting abnormalities in chest X-ray images would have high risk of being infected with COVID-19~\cite{wang2020covid}. 

We employ a large-scale of chest X-ray dataset \textit{COVIDx}~\cite{wang2020covid} to validate the proposed model. It is comprised of 18,543 chest radiography images across 13,725 cases. Example chest X-ray images belonging to normal, pneumonia, and COVID-19 classes from COVIDx dataset  are shown in Figure~\ref{ExampleImagesCOVIDX}. When we examine these examples, we can differentiate these images in terms of features shown within areas marked by the blue circle since we can observe some lighter areas indicating COVID-19 infected regions in the blue circle.

Additionally, when examining the class distribution between training and testing data, we noticed that class distribution of the training set is significantly different from that of testing set. Hence we rebuild the data by splitting the dataset into training and testing datasets that share similar class distributions, where 70\% and 30\% of data are used for training and testing datasets, respectively. The detailed information of the rebuilt dataset is shown in Table~\ref{tab1_sd} for sample distribution. 
 
\begin{table}[h!]
	\caption{Sample distribution in different classes for training and testing datasets}
        \begin{center}
                \begin{tabular}{|l|cccc|}
                \hline \textbf{Dataset} & \textbf{Normal} & \textbf{Pneumonia} & \textbf{COVID-19} & \textbf{Total} \\ \hline
                           Training		&  6,195	& 6,708	& 75  	&  12,978	 \\       
                           Testing		&  2,656	& 2,876	& 33		&   5,565	\\
                           Total		&  8,851	& 9,584	& 108	&   18,543	\\
                   \hline        
                \end{tabular}
       \end{center}
       \label{tab1_sd} 
\end{table}

We can observe that the sample distribution is extremely unbalanced regarding the number of samples of COVID-19 class. This poses a great challenge for obtaining a classifier with high performance.We overcome this challenge by the weighted cross entropy loss in the proposed model that is to assign more weight to the minority class (COVID-19 class) during the training, where the details are presented in section two. 

\subsection{Experimental settings}
In this experiment, our proposed model performed  COVID-19 classification. The key hyper parameters for training the proposed model are: 
Minibatch size: 256,
Number of epoch: 50,
Optimizer: Adam optimizer,
and Initial Learning rate: 0.1.
They are determined by trial and error. Moreover, the details of the model architecture is illustrated in Table~\ref{Table1_NetworkArchitecture}, where the residual block is the standard one~\cite{he2016deep}. Specifically, the output of the proposed model contains two parts: image class $\phi{(z^{sup})}$  and a new representation $z^{unsup}$. We employ COVID-Net\footnote{https://github.com/lindawangg/COVID-Net}~\cite{wang2020covid} as a baseline supervised model to present the state-of-the-art performance of COVID-19 image classification for comparison. Furthermore, we compared the proposed model with SRC-MT~\cite{liu2020semi} that is the state-of-the-art of semi-supervised learning since it outperformed $\Pi$ model~\cite{laine2016temporal} and mean teacher model~\cite{tarvainen2017mean}  in the area of medical image classification.

\begin{table}[h!]
	\caption{The proposed network architecture.}
	\begin{center}
	\begin{tabular}{ll}
		\hline
		\textbf{Name} & \textbf{Description}  \\ \hline
		Input &  Medical Images \\ \hline
		Shared ResNet &  one convolutional layer, 2 residual block,  \\
		&  batch normalization,  one pooling layer \\
	\hline
		Supervised ResNet &  one convolutional layer, 2 residual block,  \\
		&  batch normalization,  one pooling layer \\ \hline
		Unsupervised ResNet &  one convolutional layer, 2 residual block,  \\
		&  batch normalization,  one pooling layer \\ \hline
		Output & image class $\phi{(z^{sup})}$  and \\ 
		  &  a new representation $z^{unsup}$\\ \hline
	\end{tabular}
	\end{center}
	\label{Table1_NetworkArchitecture}
\end{table}

\subsection{Evaluation metric}

We applied different evaluation metrics to evaluate the performance of our proposed model. Since our task is a multi-class classification problem, we use accuracy, macro-average Precision (MacroP), macro-average Recall (MacroR), and macro-average Fscore (MacroF)~\cite{van2013macro, zhu2017learning, borgli2020hyperkvasir}. Accuracy is calculated by dividing the number of medical images identified correctly over the total number of testing medical images. 

\begin{equation}
	Accuracy = \frac{N_{correct}}{N_{total}}.
\end{equation}

Macro-average~\cite{yang2001study} is to calculate the metrics such as Precision, Recall and F-scores independently for each image class and then utilize the average of these metrics.  It is to evaluate the whole performance of classifying image classes.  
\begin{equation}
	MacroF = \frac{1}{C} \sum_{c=1}^{C} Fscore_c.
\end{equation}
\begin{equation}
	MacroP= \frac{1}{C} \sum_{c=1}^{C} Precision_c.
\end{equation}
\begin{equation}
	MacroR= \frac{1}{C} \sum_{c=1}^{C} Recall_c.
\end{equation}
where $C$ denotes the total number of image classes and $Fscore_c$, $Precision_c$, $Recall_c$ are $Fscore$, $Precision$, $Recall$ values in the $c^{th}$ image class which are defined by

\begin{equation}
	Fscore = \frac{2 \times Precision \times Recall}{Precision + Recall}.
\end{equation}

where $Precision$ defines the capability of a model to represent only correct image classes and $Recall$ computes the aptness to refer all corresponding correct image classes:

\begin{equation}
	Precision = \frac{TP}{TP+FP}.
\end{equation}

\begin{equation}
	Recall = \frac{TP}{TP+FN}.
\end{equation}
whereas ${TP}$ (True Positive) counts total number of medical images matched the annotated images. ${FP}$ (False Positive) measures the number of recognized classes does not match  the annotated images. ${FN}$ (False Negative) counts the number of medical images that does  not match  the predicted medical images. 
The ideal case of learning from imbalanced datasets such as \textit{COVIDx} is to improve the recall without hurting the precision. However, recall and precision goals are often conflicting, since when increasing the true positive (TP) for the minority class (True), the number of false positives (FP) can also be increased; this will reduce the precision~\cite{chawla2009data}. In addition, we employ confusion matrix to check the detailed performance for each class, especially on COVID-19 class.

\subsection{Experimental results}
\begin{table*}[ht!]
	\caption{ Comparison on supervised baseline performance.  ResNet is trained on different ratios (\%) of labeled X-ray images.  Weighted ResNet is built by assigning more weight to COVID-19 class during training for overcoming the challenge of data imbalance. }
       
        \begin{center}
                \begin{tabular}{|l|cccc|}
                \hline \textbf{DL} & \textbf{Accuracy (\%)} & \textbf{MacroP (\%)} & \textbf{MacroR  (\%)} & \textbf{MacroF  (\%)} \\ \hline 
                		  COVID-Net (100\%)	&  93.98		& 72.70 		& 96.05		&  77.33 	\\ 
		  \textbf{ResNet (100\%)}		&   \textbf{94.68}	& \textbf{90.52}		& \textbf{78.87}		& \textbf{83.04}	\\
		  \hline 
                	           ResNet (5\%)			&  87.59 		& 58.23 		& 58.50 		&  58.30  	\\
                 	   ResNet (7\%)			&  89.27 		& 59.51 		& 59.91 		&  59.68  	\\
			   ResNet (9\%)			&  90.10 		& 60.05  		& 60.44 		&  60.24  	 \\
			   \hline 			   
			   Weighted ResNet (5\%)		&  86.45 		& 57.63 		& 57.95 		&  57.78  	\\
                 	   Weighted ResNet (7\%)		&  88.70 		& 59.15 		& 59.54 		&  59.31  \\
			   Weighted ResNet (9\%)		&  89.36 		& 66.16  		& 60.46  		&  60.78   \\
                    	   %ResNet (10\%)		&  88.83 		& 59.23 		& 59.53		&  59.37  	\\
			   %ResNet (20\%)		&  92.30 		& 94.85  		& 66.12  		&  69.19  	 \\
			    \hline
	   		       
                     \hline
                      
                \end{tabular}
       \end{center}
         \label{tab3_covid-19}
\end{table*}

\newcommand{\photo}[1]
{
    \includegraphics[width=5.5cm]{#1}
}

\begin{figure*}[h!]
 \begin{center}
\begin{tabular}{cc}
	\photo{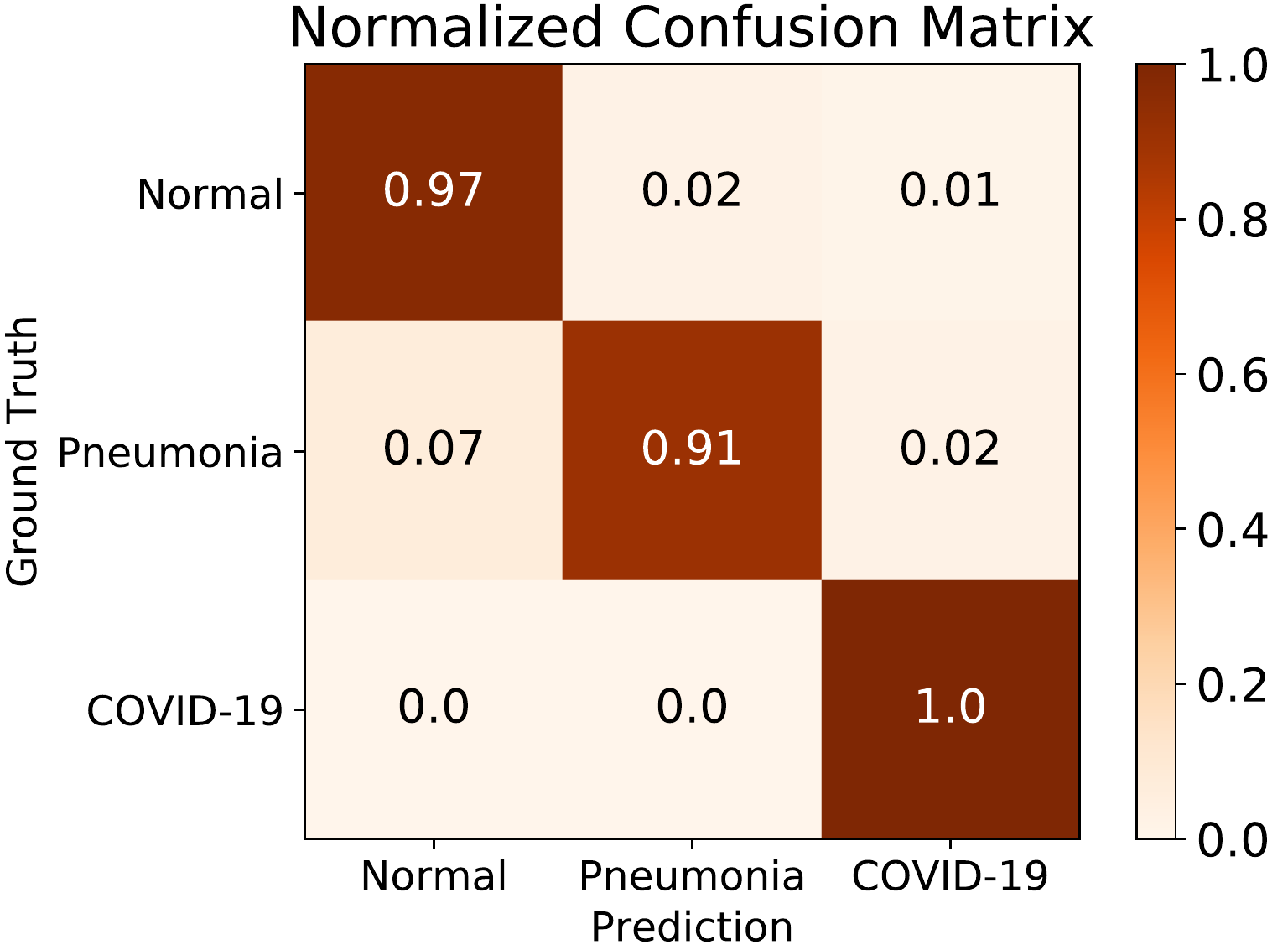} &  \photo{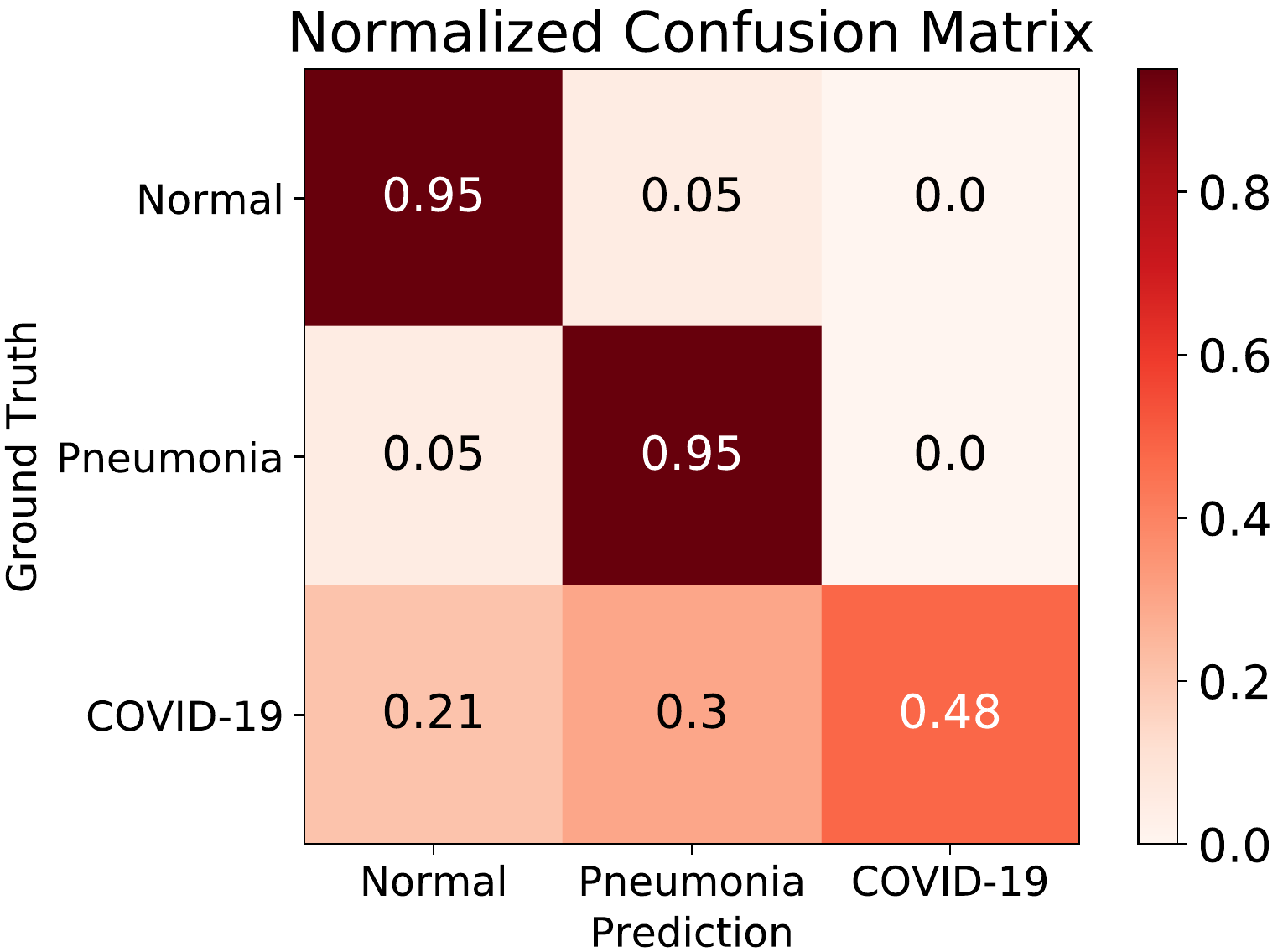} \\
	(a) COVID-Net (100\%) &  (b) ResNet (100\%)  \\	
\end{tabular}
 \end{center}
 \caption{Comparison of confusion matrix generated by COVID-Net (100\%) and ResNet (100\%).}
 \label{Fig_covid-net}
\end{figure*}

We evaluated the proposed model performance in four steps. The first step is to examine the performance of supervised learning baselines, which is to prove if ResNet is a reasonable supervised model for COVID-19 image classification. A competitive supervised baseline is useful to compare the proposed semi-supervised model in order to present the effectiveness of the proposed model. Furthermore, we will check whether fewer labeled data will lead to lower performance. The second step is to comparing the proposed model with  state-of-the-art semi-supervised learning. The third step is to examine whether the hyper-parameter setting will affect the performance of the proposed model significantly. Finally, we will discuss why the proposed model cannot classify certain COVID-19 cases.

\subsubsection{Supervised learning for COVID-19 classification}

\begin{figure*}[h!]
 \begin{center}
\begin{tabular}{ccc}
	\photo{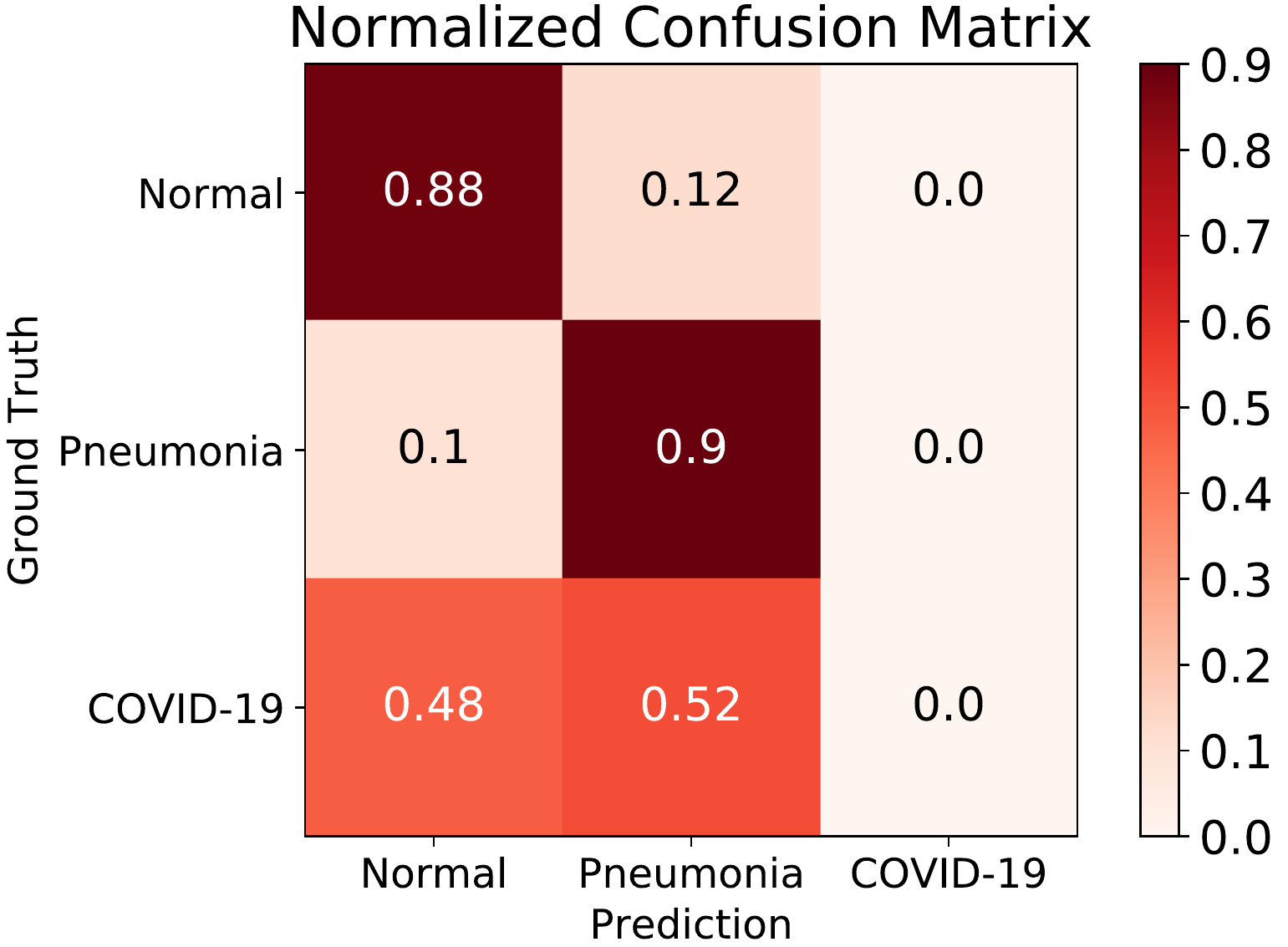} &  \photo{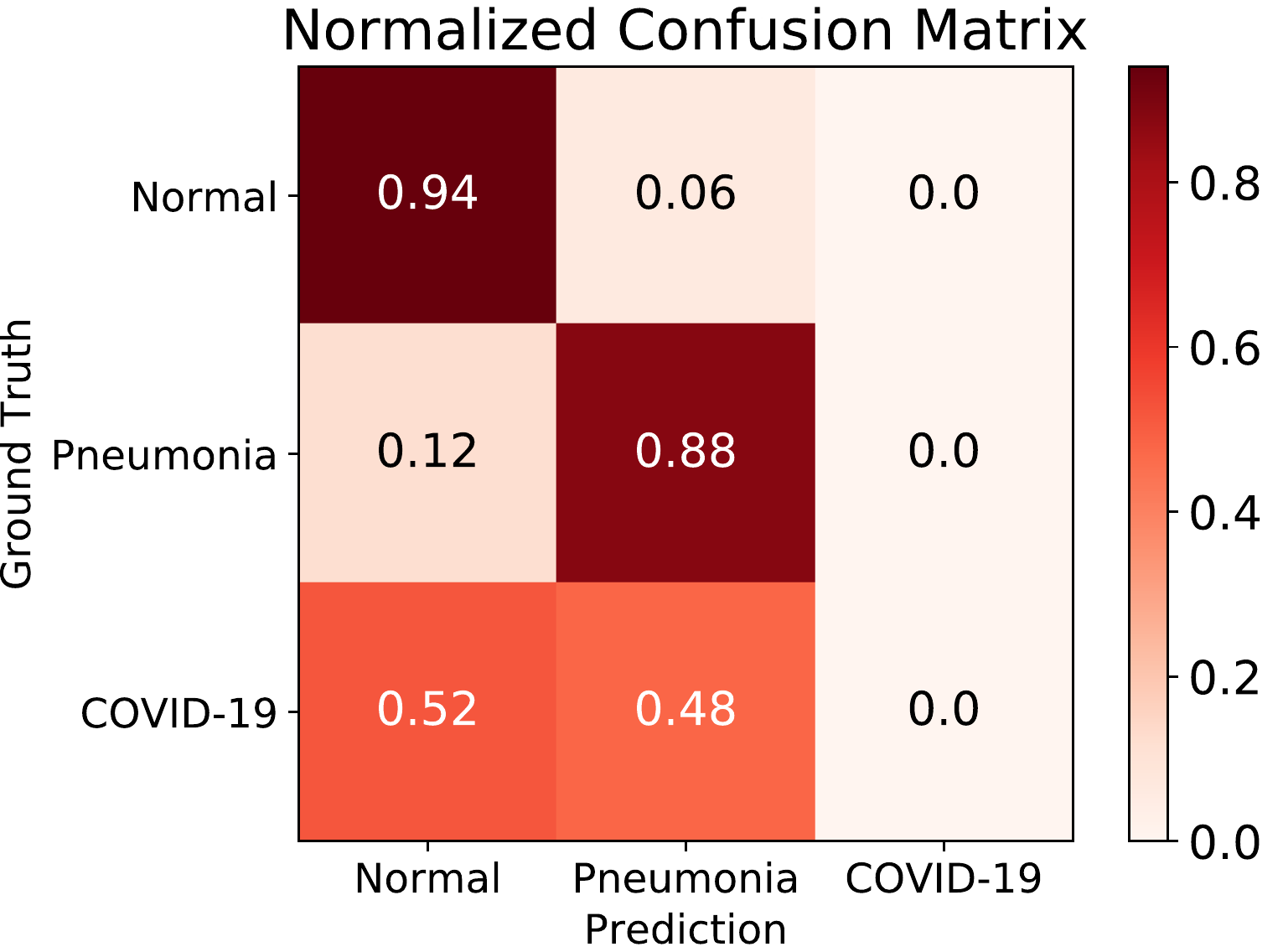} &  \photo{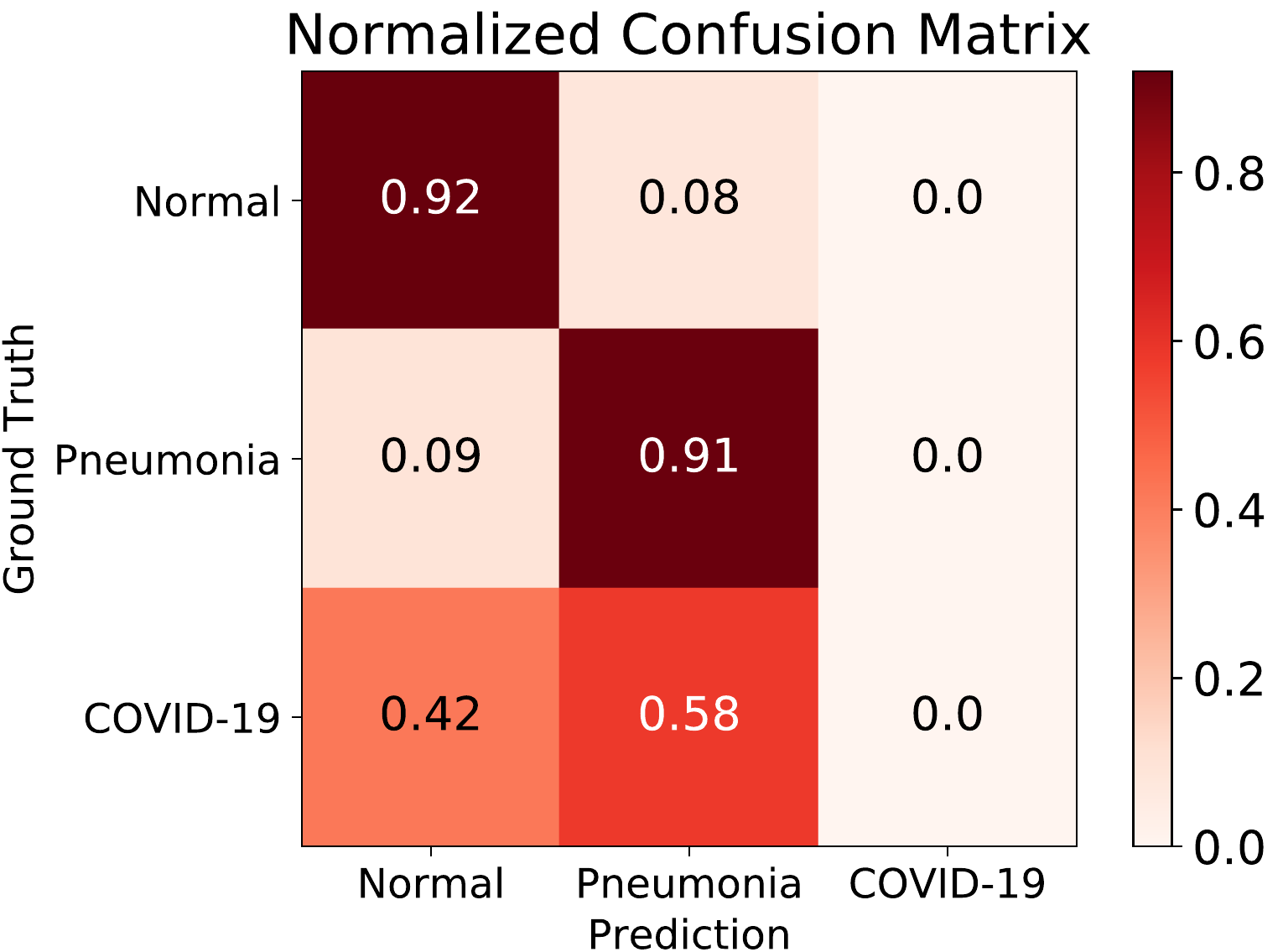}\\
	(a) ResNet (5\%) &  (b) ResNet (7\%)  &  (c) ResNet (9\%) \\
	\photo{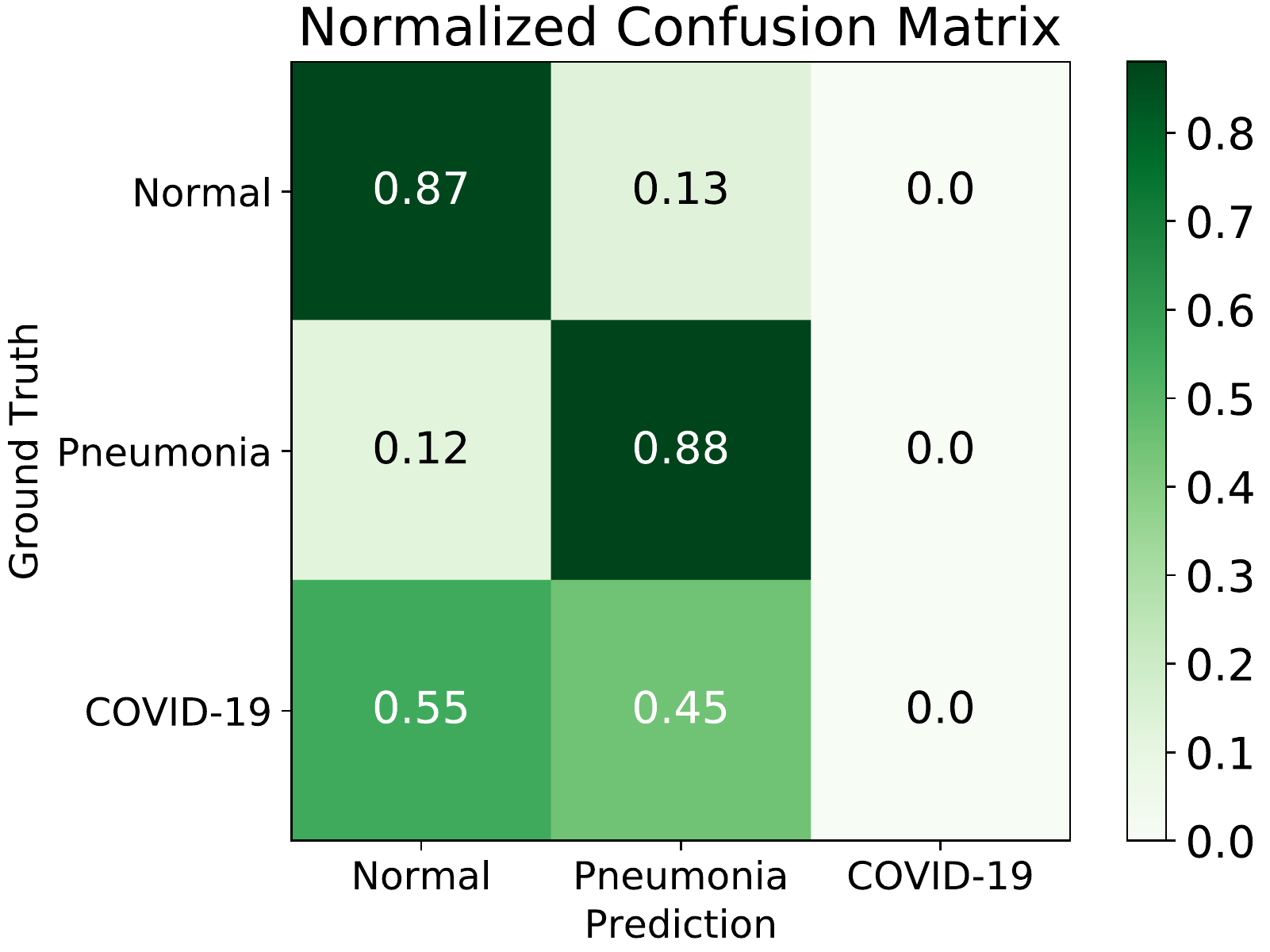} &  \photo{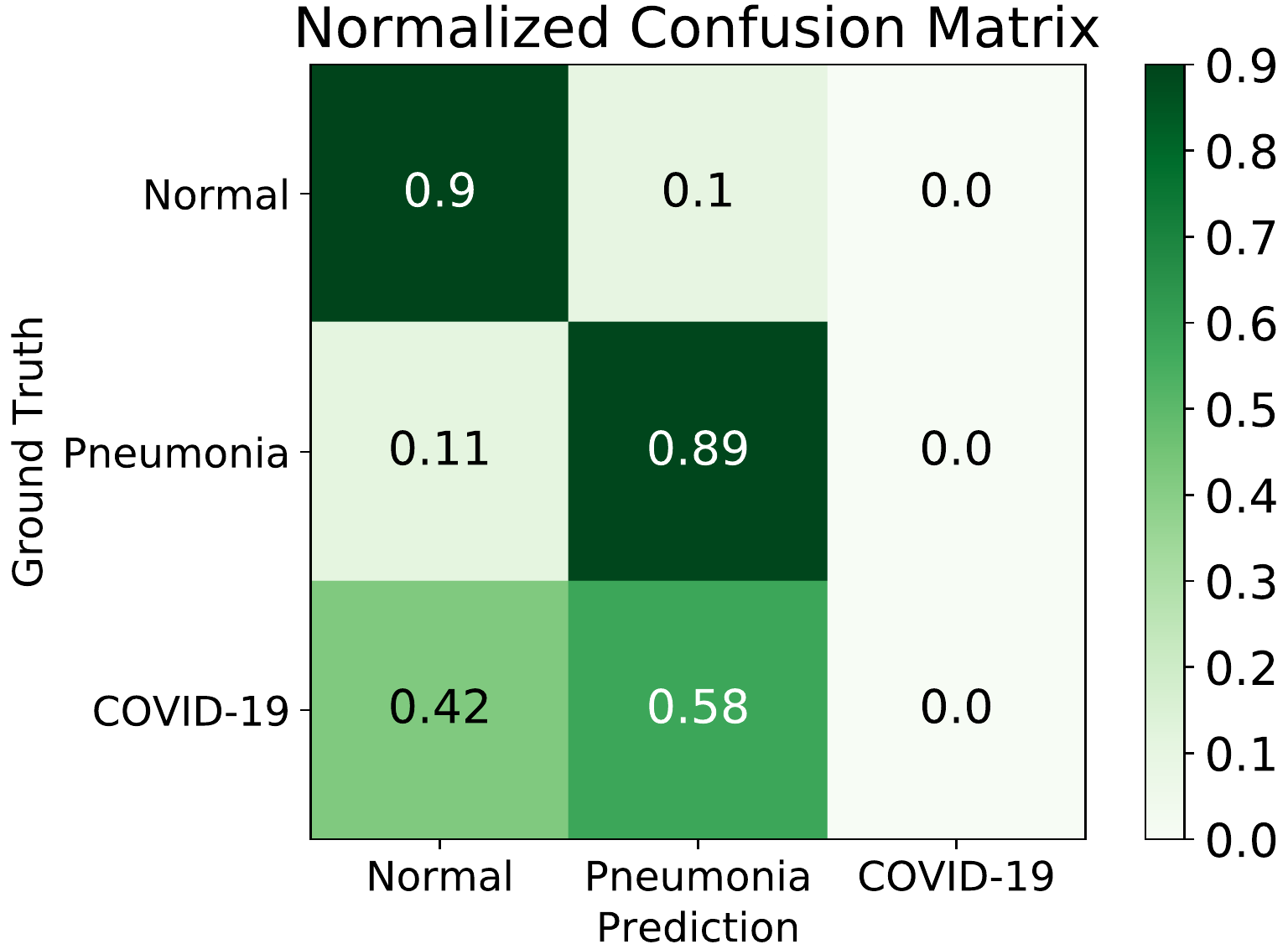} &  \photo{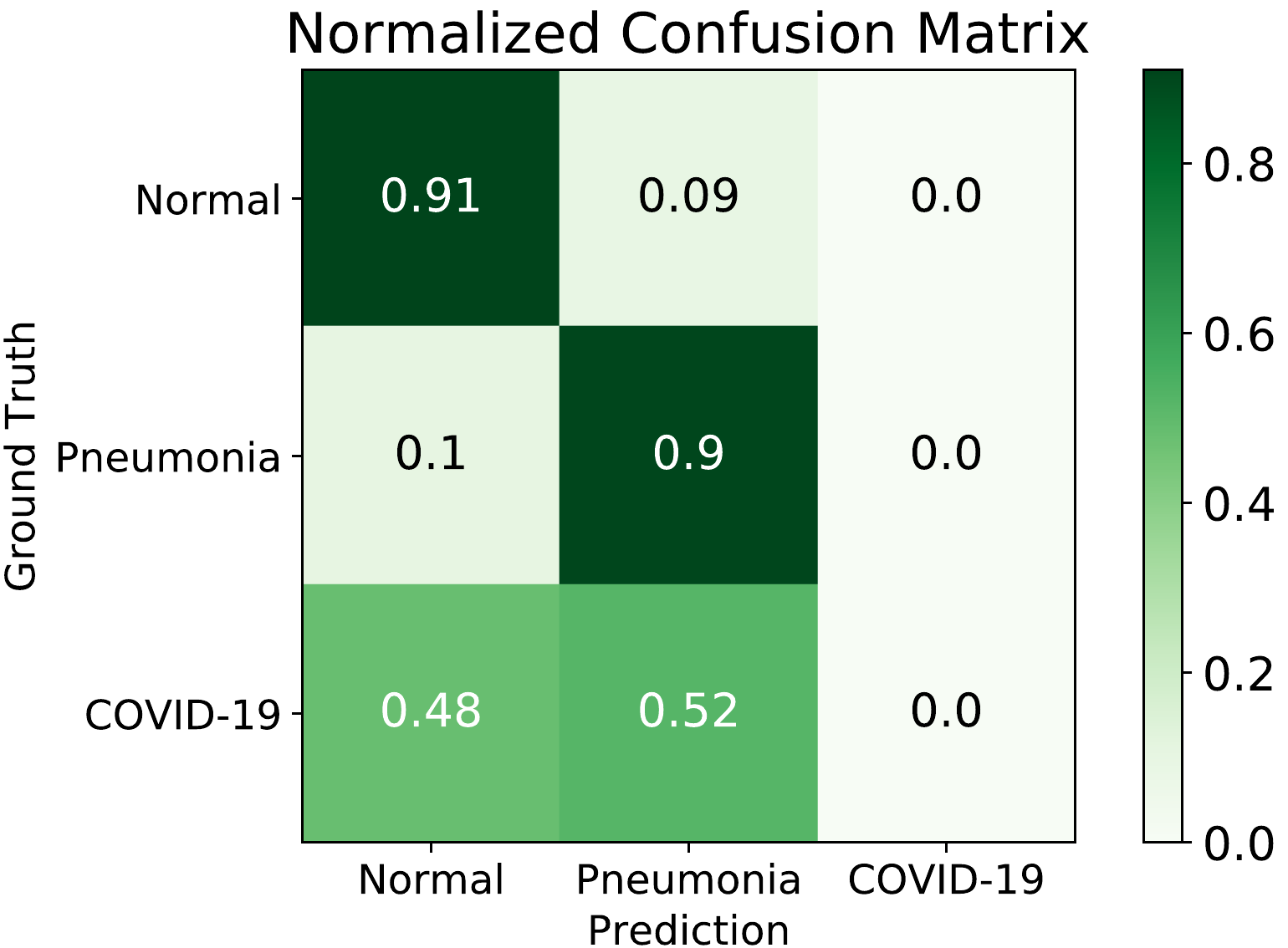}\\
	(d) Weighted ResNet (5\%) &  (e) Weighted ResNet (7\%)  &  (f) Weighted ResNet (9\%) \\
	
\end{tabular}
 \end{center}
 \caption{Comparison of confusion matrix generated by different ResNets training on different ratios of labeled data. Weighted ResNet is built by assigning more weight to COVID-19 class during training for overcome the challenge of data imbalance.}
 \label{Fig3_covid-19}
\end{figure*}

Table~\ref{tab3_covid-19} presents the comparison of supervised baselines built with ResNet. We observed that ResNet (100\%) can outperformed COVID-Net (100\%) when comparing accuracy, macro-average precision, and macro-average Fscore. It means that ResNet is a competitive supervised baseline for COVID-19 image classification. Additionally, for learning on fewer labeled data, we only focus on the cases of 5\%, 7\%, and 9\% labeled data since the labeled data will be very scarce in medical domain~\cite{litjens2017survey} during the early stage of a global pandemic such as COVID-19 outbreak.  We observed that the classification accuracy can be improved by increasing the labeled data to train ResNet. Meanwhile, the performance such as accuracy and MacroF is reduced significantly when comparing with ResNet (100\%), which demonstrates that more labeled data is imperative for building high-performance supervised models. Moreover, we observed that weighted ResNet cannot improve the performance since we might assign inappropriate weight to different classes.

On the other side, we compare their confusion matrix to examine the performance details in Fig.~\ref{Fig_covid-net}. It indicates that ResNet (100\%) can be a promising supervised baseline model when compared to COVID-Net in terms of the accuracies on the normal and pneumonia classes. For the COVID-19 class, ResNet is lower than COVID-Net since COVID-Net employed transfer learning to enhance performance. 

To check the performance for each class when learning on fewer labeled data, we present the detailed performance with confusion matrix shown in Fig.~\ref{Fig3_covid-19}. When we use low ratios of labeled training data to train models, ResNet cannot recognize COVID-19 images effectively, which is duo to insufficient COVID-19 labeled samples. In the training sets of these cases, only few of images are for the COVID-19 class. For example, in the case of ResNet (5\%), we only have three images for COVID-19 class in the training data, which means most of training images are for the classes of Normal and Pneumonia. Learning on this data will lead to classification bias. Weighted ResNet was no sufficient to enhance the performance, which means even more weight assigned to COVID-19 class is not enough to overcome the lack of labeled samples to learn distinguish features to differentiate COVID-19 patients from Non-COVID-19 patients on X-ray images with supervised learning.

\subsubsection{Comparing the proposed model with state-of-the-art semi-supervised learning}

\begin{table*}[h!]
	\caption{\label{tab1_ml} Comparing performance between SRC-MT and Our model (Semi-supervised ResNet (SSResNet)).}
       
        \begin{center}
                \begin{tabular}{|l|cccc|}
                                 \hline \textbf{Semi-supervised Model} & \textbf{Accuracy (\%)} & \textbf{MacroP (\%)} & \textbf{MacroR  (\%)} & \textbf{MacroF  (\%)} \\ \hline 
                 	   SRC-MT (5\%)		&  90.67 		& 61.08 		& 60.75 		&  60.59  	\\
                 	   SRC-MT (7\%)		&  89.82 		& 89.92 		& 74.13 		&  78.95  	\\
			   SRC-MT (9\%)		&  92.79 		& 93.61  		& 79.15  		&  84.15  \\
\hline 			   

                    \hline \textbf{Our model} & \textbf{Accuracy (\%)} & \textbf{MacroP (\%)} & \textbf{MacroR  (\%)} & \textbf{MacroF  (\%)} \\ \hline
                    	   SSResNet (5\%)	&  84.95 		& 61.18 		& 66.76 		&  62.41  	\\
	   		   SSResNet (7\%)	&  84.21 		& 63.67 		& 67.85 		&  62.83  \\
			   SSResNet (9\%)	&  81.79 		& 59.34 		& 70.99 		&  59.19  	\\	   	\hline  
                \end{tabular}
       \end{center}
         \label{tab4_covid-19}
\end{table*}
\begin{figure*}[h!]
	\begin{center}
	\begin{tabular}{ccc}
	\photo{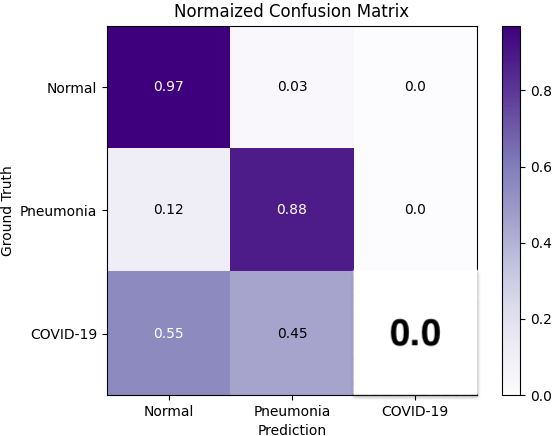} &  \photo{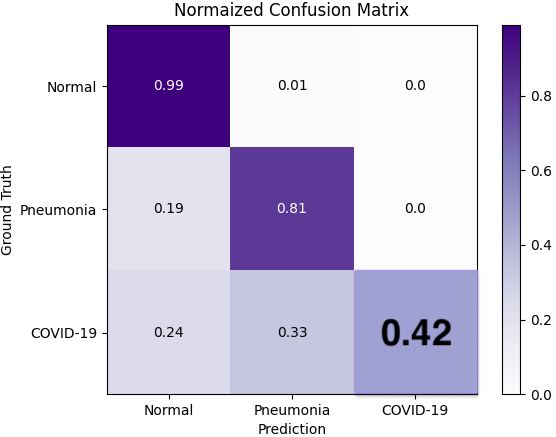} &  \photo{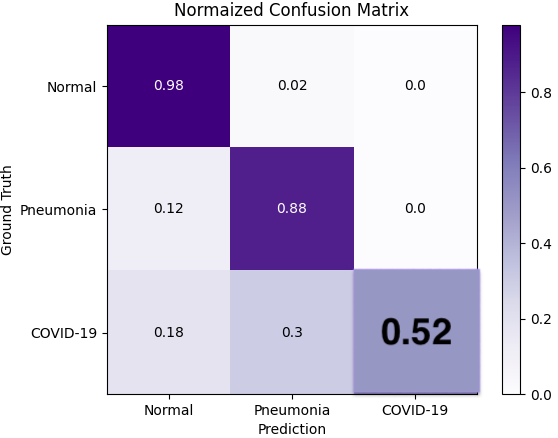}\\
	(a) SRC-MT (5\%) &  (b) SRC-MT (7\%)  &  (c) SRC-MT (9\%) \\	
	\photo{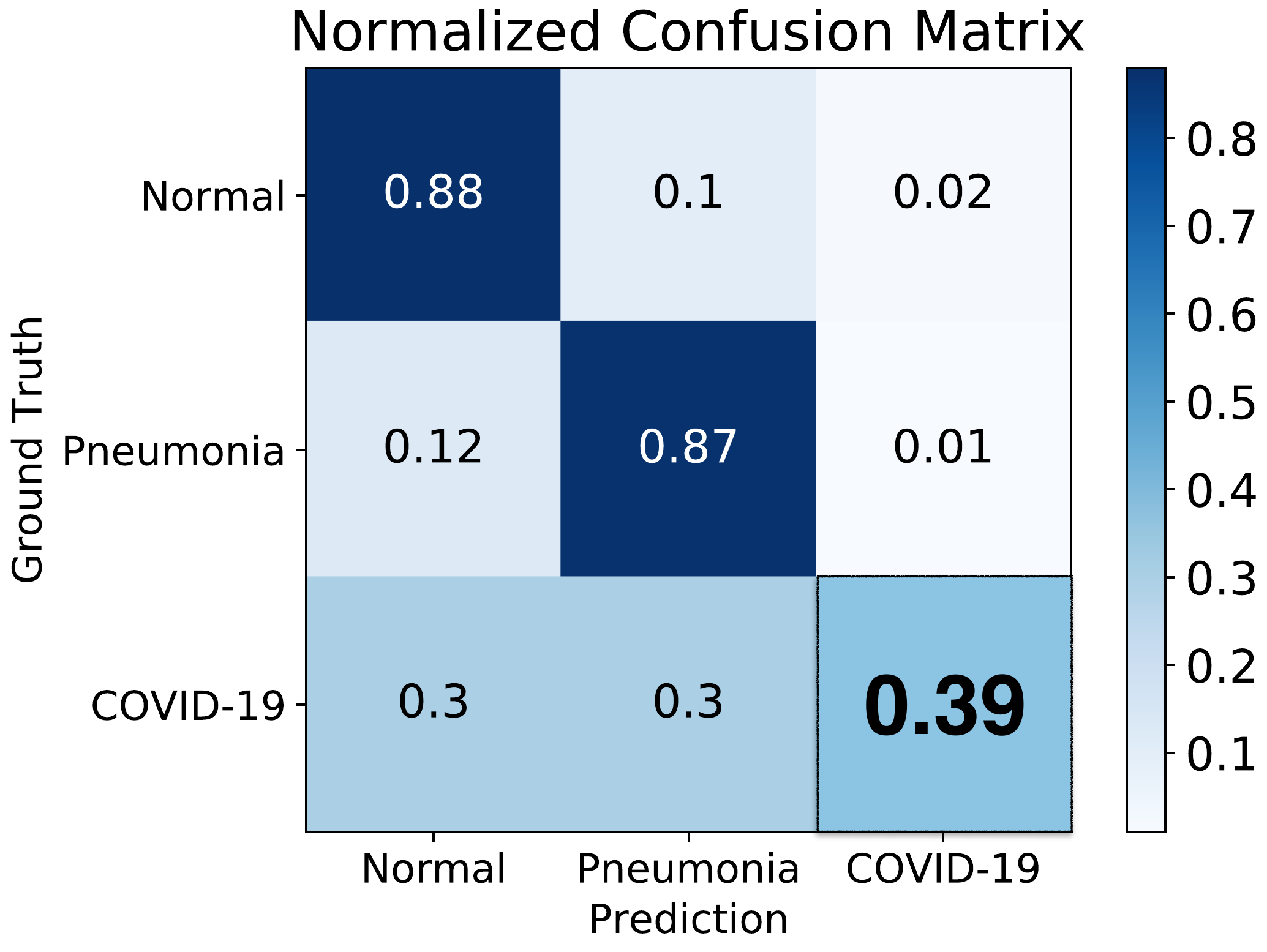} &  \photo{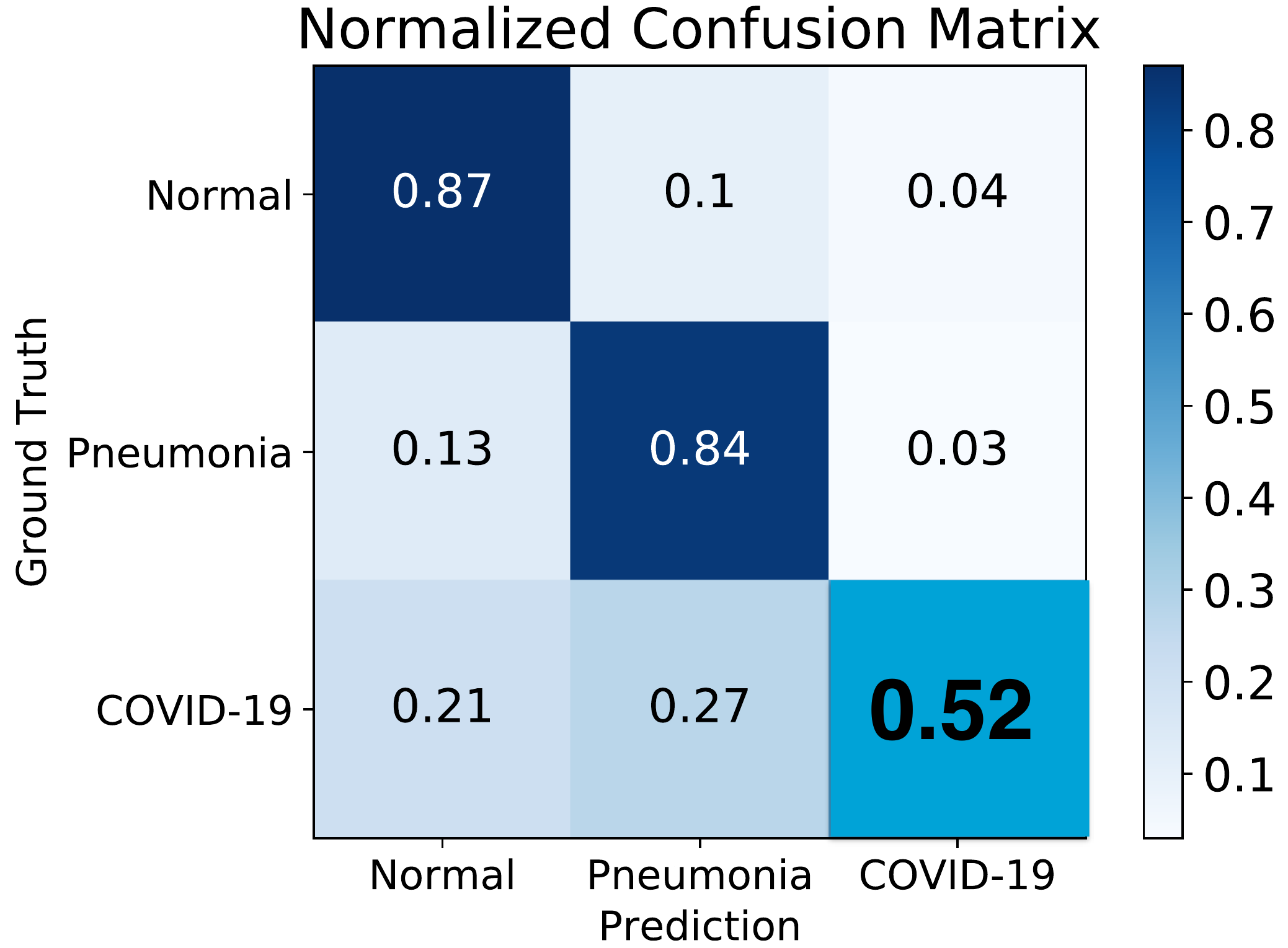} &  \photo{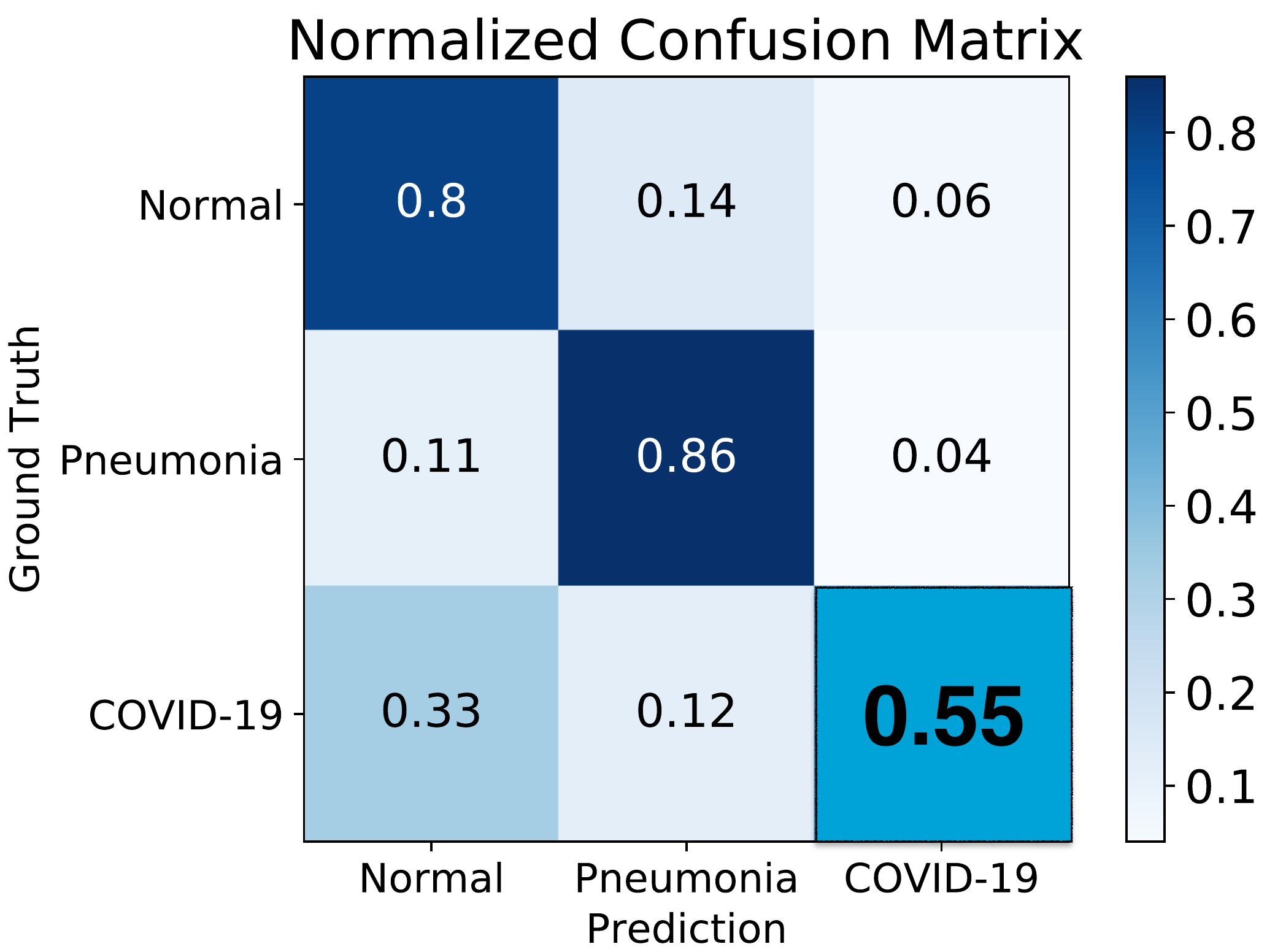}\\
	(d) SSResNet (5\%) &  (e) SSResNet (7\%)  &  (f) SSResNet (9\%) \\	
	\end{tabular}
 	\end{center}
 \caption{Comparison of confusion matrix generated with SRC-MT and SSResNets  trained on different ratios of labeled data. }
 \label{Fig4_covid-19}
\end{figure*}

In this section, we will examine if the proposed model is able to effectively identify COVID-19 samples by training on very limited amount of annotated images. Table~\ref{tab4_covid-19} presents the comparison of classification performance between SRC-MT and the proposed model (SSResNet). Overall accuracies of SRC-MT are better than those of the proposed model. However, when only 5\% labeled samples were used for training, MacroF of our proposed model is higher than that of SRC-MT, which indicates that the proposed model is more effective in detecting COVID-19 samples. can detect COVID-19 samples with higher performance. It means  that compared to SRC-MT, the unsupervised path could enhance the data representation for improving COVID-19 classification more effectively. 

In addition, we examine detailed performance of each class with confusion metrics shown in  Fig.~\ref{Fig4_covid-19}. We observed that the accuracy of recognizing COVID-19 by the proposed model is higher than that of SMC-TC, which means SSResNets can learn more effective features from unlabeled data to recognize COVID-19 samples. Furthermore, with the increased ratios of labeled data, the accuracies of recognizing COVID-19 is enhanced significantly. It means that the unsupervised path can enhance the representations of images to improve the classification. In other words, unlabeled data contributed to increasing the COVID-19 classification performance significantly by enhancing the image representations with the unsupervised path of the SSResNet.

\begin{table*}[h!]
	\caption{\label{tab_label_weight} Comparing performance with different class weights. $c_{1}$, $c_{2}$, and $c_{3}$  are the weights of Normal class, Pneumonia class, and COVID-19 class, respectively.} 
        \begin{center}
                \begin{tabular}{|c|cccc|}
                    \hline  &  & 5\% Labeled Data &  & \\ 
                    \hline \textbf{Class Weight} & \textbf{Accuracy} & \textbf{MacroP} & \textbf{MacroR} & \textbf{MacroF}\\ \hline
                    	   $c_{1}$ = 1, $c_{2}$ = 1, $c_{3}$ = 2			&  84.95 		& 61.18 		& 66.76 		&  62.41  	\\
                            $c_{1}$ = 1, $c_{2}$ = 1, $c_{3}$ = 5			&  78.82 		& 58.18 		& 67.38 		&  56.79  	\\  
                            $c_{1}$ = 1, $c_{2}$ = 1, $c_{3}$ = 10			&  66.78 		& 57.24 		& 66.70 		&  50.77  	\\                   \hline
                    \hline  &  & 7\% Labeled Data &  & \\
                    \hline \textbf{Class Weight} & \textbf{Accuracy} & \textbf{MacroP} & \textbf{MacroR} & \textbf{MacroF}\\ \hline
                    	   $c_{1}$ = 1, $c_{2}$ = 1, $c_{3}$ = 2			&  85.90 		& 65.75 		& 64.62		&  64.84  \\
                            $c_{1}$ = 1, $c_{2}$ = 1, $c_{3}$ = 5			&  84.21 		& 63.67 		& 67.85 		&  62.83  	\\ 
                            $c_{1}$ = 1, $c_{2}$ = 1, $c_{3}$ = 10			&  79.31 		& 58.57 		& 67.71 		&  59.39 	\\
                      \hline
                     \hline  &  & 9\% Labeled Data &  & \\ 
                      \hline \textbf{Class Weight} & \textbf{Accuracy} & \textbf{MacroP} & \textbf{MacroR} & \textbf{MacroF}\\ \hline
                    	   $c_{1}$ = 1, $c_{2} = $1, $c_{3}$ = 2			&  87.28 		& 70.32 		& 62.93 		&  64.65  	\\
                            $c_{1}$ = 1, $c_{2}$ = 1, $c_{3}$ = 5			&  84.69 		& 60.91 		& 66.79 		&  61.86  	\\   
                            $c_{1}$ = 1, $c_{2}$ = 1, $c_{3}$ = 10			&  81.79		& 59.34 		& 70.99 		&  59.19  	\\             \hline
                    
                \end{tabular}
       \end{center}
         \label{tab_batch_size}
\end{table*}

\begin{figure*}[t] 
	\centering
	%\captionsetup{font={footnotesize }}
	\subfigure[Normal]{
		\label{Fig-01-1}
		\includegraphics[height=1.85in]{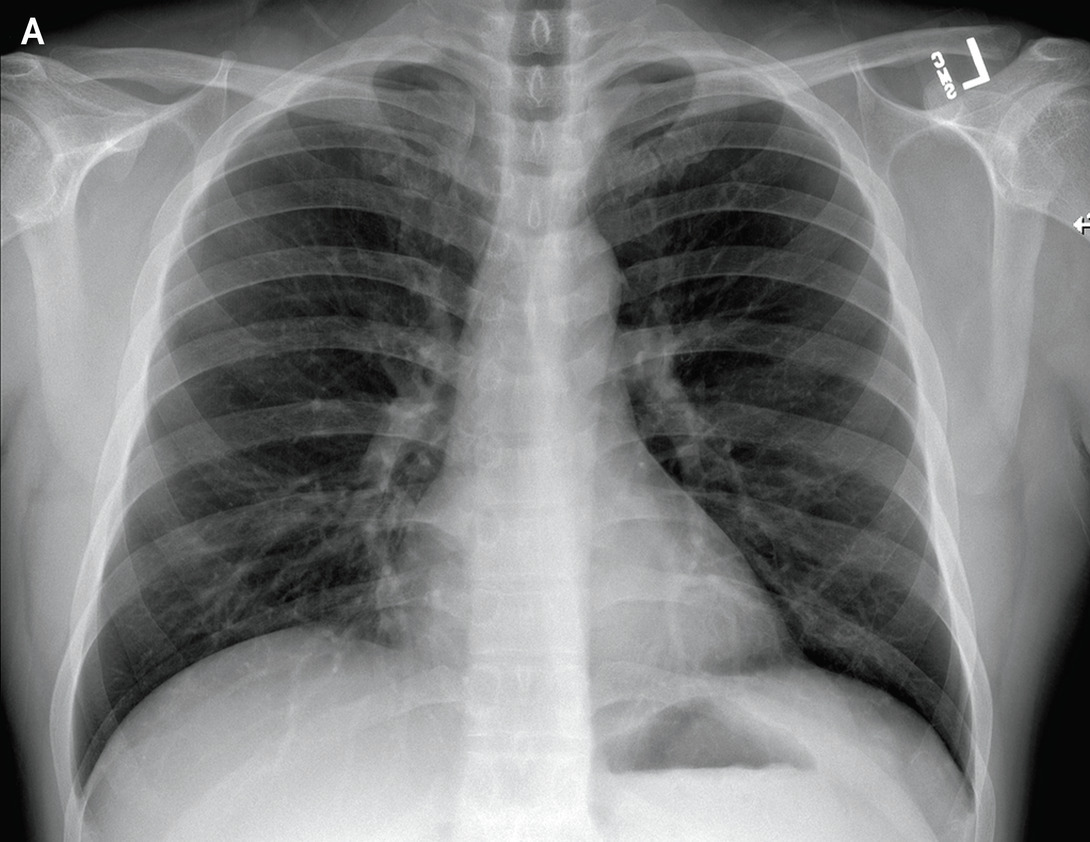}} 
	\hspace{0.02in}
	\subfigure[COVID-19]{
		\label{Fig-01-2}
		\includegraphics[height=1.85in]{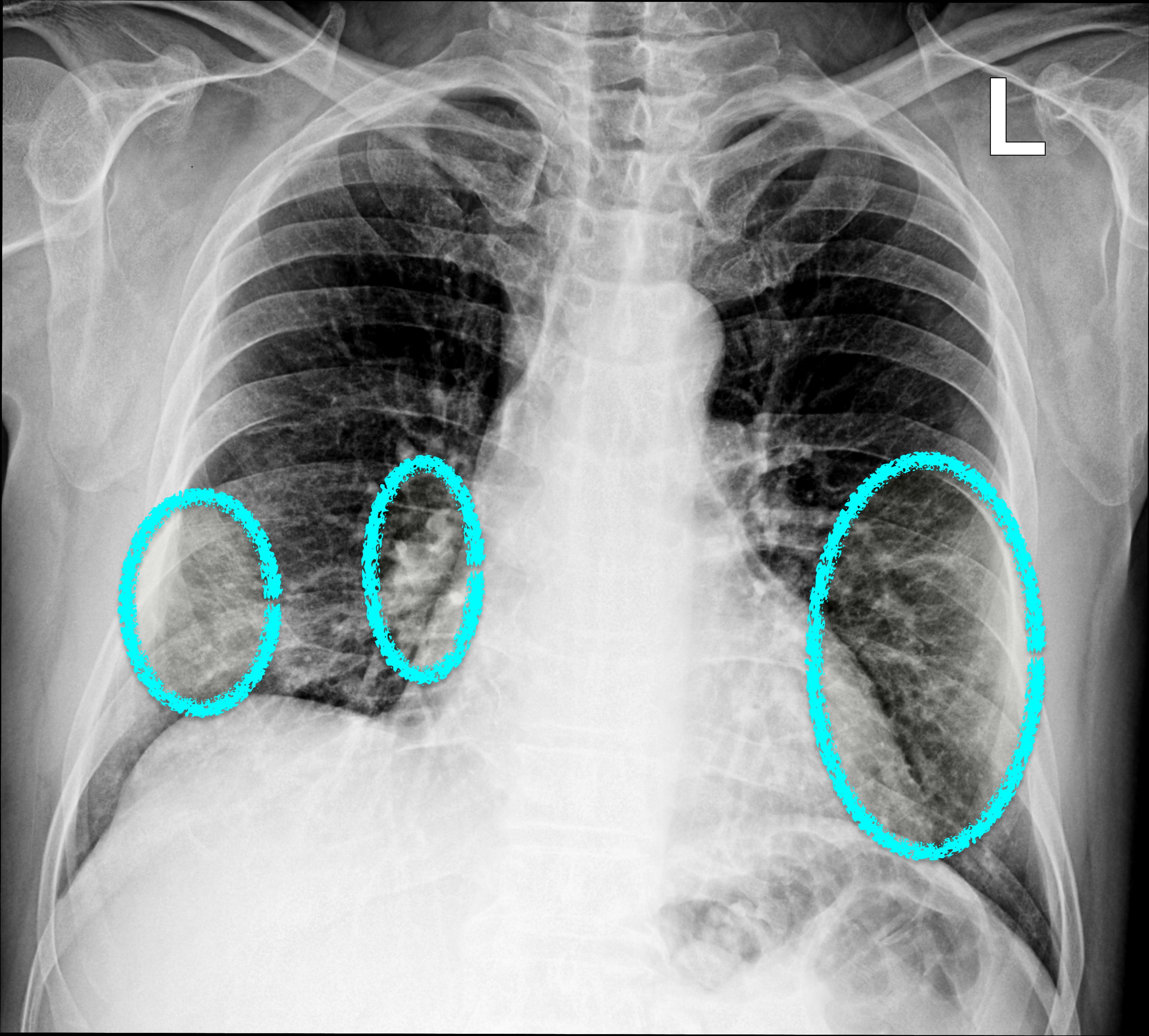}}
	\hspace{0.02in}
	\subfigure[Pneumonia]{
		\label{Fig-01-3}
		\includegraphics[height=1.85in]{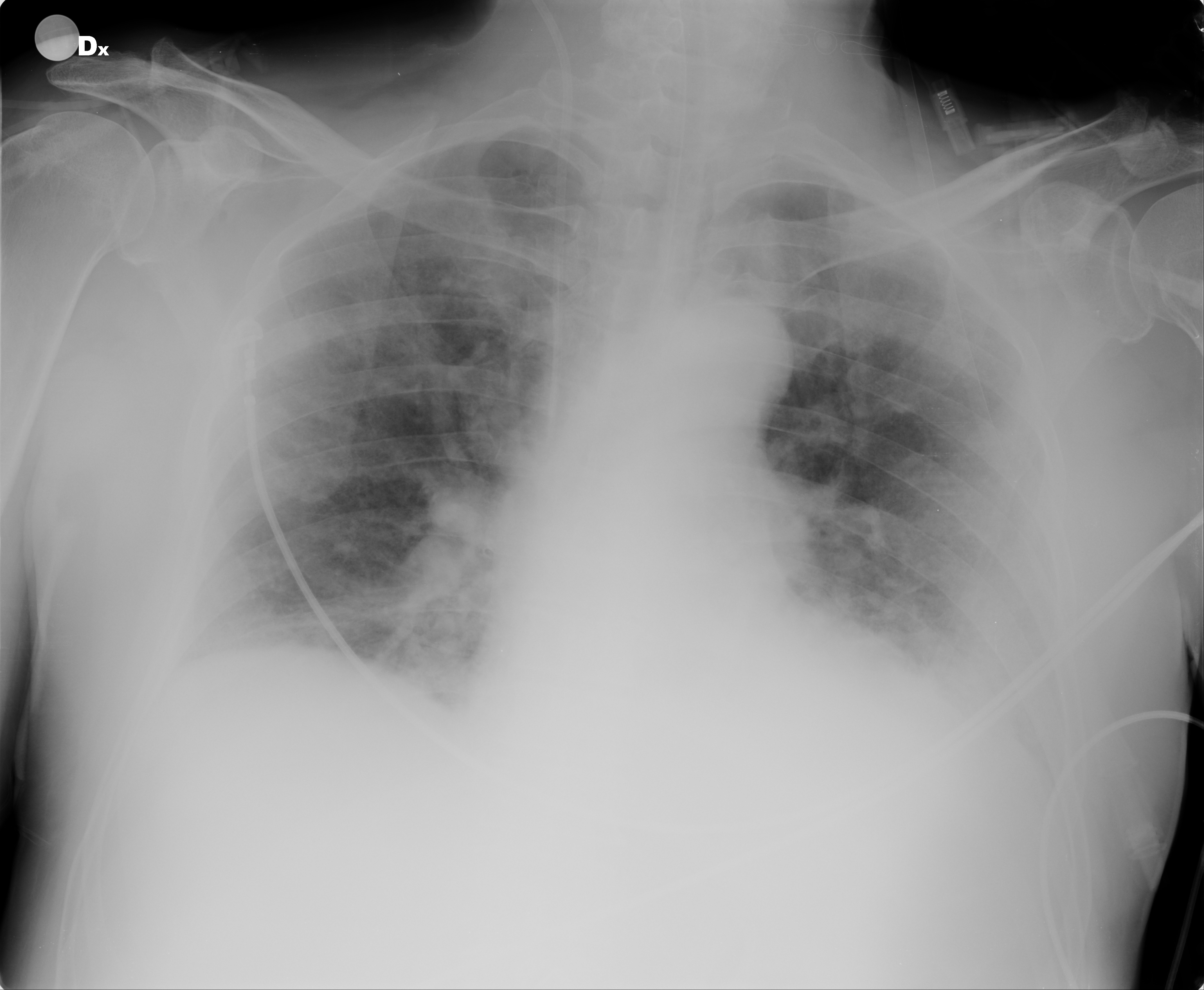}}
	\hspace{0.02in}
	\caption{COVID-19 samples classified into Normal, COVID-19, and Pneumonia classes, are shown in (a), (b), and (c) respectively.  The blue circles locate the infected regions of COVID-19. }
	\label{ExampleError}
\end{figure*}

\subsubsection{Hyper-parameter setting}

In addition to examining the performance comparison between the proposed models and baselines, we have to figure out whether the proposed model is sensitive to the hyper-parameters.  There are various hyper-parameters involved in the learning procedure of the proposed model. Here, we choose class weight to check since different weights would lead to different performance of recognizing COVID-19 samples. Table~\ref{tab_label_weight} shows the comparison results for different weights of three classes. We observe that different weights will result in significant differences of the performance when examining  the values of accuracy. On the other hand, compared to accuracy and MacroP,  MacroR and MacroF are less sensitive to the weight of COVID-19 class. Generally, we have to delicately select the weight for COVID-19 class  to obtain the optimal performance.

\subsubsection{Error Analysis}

Fig.~\ref{ExampleError} presents three COVID-19 samples that are classified into Normal, COVID-19, and Pneumonia classes, respectively. X-ray images of COVID-19 patients shows various features for different stages of COVID-19 patients\footnote{https://www.uclahealth.org\/radiology\/covid-19-chest-x-ray-guideline}. At the early stage of COVID-19 patients, X-ray images cannot present significant features (Fig.~\ref{ExampleError} (a)) that can be used to differentiate COVID-19 and Non COVID-19 patients, which leads to the incorrect classification result for the sample. It is consistent with the expectation that X-ray images are not ideal evidences to support diagnosis of COVID-19 for the patients at the early stage. 

However, with development of COVID-19, X-ray images are able to present obvious features such as multifocal lung airspace opacities, nodules  and consolidation (Fig.~\ref{ExampleError} (b)), which contributes to the correct classification result. Unfortunately, if the patients are at the late stage of COVID-19, X-ray images presents lobar diffused consolidation (See Fig.~\ref{ExampleError} (c)) that is similar to features of pneumonia. These features will be confusing to the proposed model and lead to the incorrect result  for the sample shown in Fig.~\ref{ExampleError} (c). In summary, in terms of samples shown in Fig.~\ref{ExampleError}, the proposed model will be effective for the patients who are in the development of COVID-19 rather than those at the early stage or late stage of such disease.

\section{Related Work}
\label{sec4}

Deep learning technique has shown its power on classification of COVID-19. Ghoshal \textit{et al.}~\cite{ghoshal2020estimating} proposed a Bayesian convolutional neural network to estimate the diagnosis uncertainty in COVID-19 prediction, where the dataset includes 70 lung X-ray images of patients with COVID-19 from an online COVID-19 dataset~\cite{cohen2020covid}, and non-COVID-19 images from Kaggle’s Chest X-Ray data (Pneumonia). Narin \textit{et al.}~\cite{narin2020automatic} is to detect COVID-19 infection from X-ray images through comparing three different deep learning models, namely, ResNet50, InceptionV3, and Inception-ResNetV2.  The evaluation results show that the ResNet50 model outperformed other two models.
Zhang \textit{et al.}~\cite{zhang2020covid} also utilized ResNet to complete COVID-19 classification on X-ray images and estimated an anomaly score to optimize the COVID-19 score for the classification. In addition, Wang \textit{et al.}~\cite{wang2020covid} propose COVID-Net to detect COVID-19 cases using X-ray images. In general, most current studies use X-ray images to differentiate between COVID-19 and other pneumonia and healthy subjects. 
%However, with limited amount of COVID-19 images, it is insufficient to evaluate the robustness of the methods and also poses questions to the generalizability. 

In addition to COVID-19 image classification, it is imperative to figure out the regions of infection of COVID-19  since it will provide detailed information on COVID-9 for diagnosis. Semantic segmentation is able to help us recognize the regions and corresponding patterns to assess and quantify COVID-19, where the regions of interest (ROIs) contains those of lung, lobes, bronchopulmonary segments, and infected regions or lesions, in the chest X-ray or CT images. Moreover, segmented regions could be further used to extract handcrafted or self-learned features for diagnosis and other applications. Deep learning has promoted the development of semantic segmentation of images significantly~\cite{ronneberger2015u, cciccek20163d}.  To segment ROIs in CT, the segmentation networks for COVID-19 include classic U-Net~\cite{zheng2020deep, huang2020serial, gozes2020rapid}, UNet++ ~\cite{chen2020deep}, and VB-Net~\cite{shan2020lung}. The segmentation methods related to COVID-19  can be classified into two groups: 1) the lung-region-oriented methods and 2) the lung-lesion-oriented methods. The first group aims at separating lung regions, i.e., whole lung and lung lobes, from other (background) regions in CT or X-ray images~\cite{jin2020ai, tang2020severe}. For example, Jin \textit{et al.}~\cite{jin2020ai} is to detect the whole lung region with UNet++. The second group is to detect lesions (or metal and motion artifacts) in the lung from lung regions~\cite{cao2020longitudinal, gaal2020attention}. The experimental results indicate that the segmentation of X-ray images is even more challenging because of the ribs projected onto soft tissues in 2D.  Although supervised deep learning outperforms other models on these two tasks, it requires substantial amount of labeled data to train the model, which is not practical in real applications.

Semi-supervised deep learning has attracted lots of attentions since it has the strong ability to generalize the model performance through learning on labeled data and unlabeled data~\cite{weston2008deep, kingma2014semi, rasmus2015semi, laine2016temporal}. Generally, it is to train the deep neural networks by jointly optimizing the standard supervised classification loss on labeled samples and an unsupervised loss on unlabelled data~\cite{weston2008deep, laine2016temporal}. The rationale of these semi-supervised learning models is to enrich the supervision signals by exploiting the knowledge learned on unlabeled data~\cite{lee2013pseudo}, or regularize the network by enforcing smooth and consistent classification boundaries~\cite{rasmus2015semi}. Regarding COVID-19 research such as COVID-19 image classification and image segmentation, semi-supervised learning is employed to resolve the lacking of labeled data~\cite{zhou2020soda, calderon2020correcting, paticchio2020semi, ma2020active, yang2020federated, berenguer2020explainable}. However, for COVID-19 image classification, these studies~\cite{zhou2020soda, calderon2020correcting, paticchio2020semi} have not comprehensively examined the model performance on a large-scale of X-ray image dataset such as \textit{COVIDx}~\cite{wang2020covid} by comparing with the state-of-the-art, especially for the case of very few labeled data such as less than 10\% labeled data. This paper proposed a semi-supervised deep learning model for COVID-19 image classification and checked out the model performance systematically on the \textit{COVIDx}~\cite{wang2020covid} dataset.

\section{Conclusion and Future Work}
\label{sec5}

In this paper, a novel framework of semi-supervised deep learning is proposed for COVID-19 image classification on chest X-ray images. Supervised learning based COVID-19 classification on X-ray datasets could provide useful information to medical staff for facilitating a diagnosis of COVID-19 in an effective and efficient manner.  Unfortunately, it relies on the availability of large amount of labeled medical images, which are not available in practice in the early outbreak of such global pandemic. Hence, 
we propose a semi-supervised learning model based on ResNet that can utilize unlabeled images to enhance classification performance.  There are two paths in the model for reducing  supervised cross entropy loss  and unsupervised mean squared error loss, respectively. Then training is performed by jointly optimizing these two losses, which allows the proposed scheme to take advantage of the information from both labeled and unlabeled images. Experimental results demonstrate that the proposed model could recognize COVID-19 lung pathology effectively by learning on very limited labeled images and substantial unlabeled images. For the future work, we plan to extend the proposed model for other tasks such as COVID-19 image segmentation.

\appendices

\section*{Acknowledgment}
\label{acknowledgement}
This research work is supported in part by the Texas A\&M Chancellor's Research Initiative (CRI), the U.S. National Science Foundation (NSF) award 1464387 and 1736196, and by the U.S. Office of the Under Secretary of Defense for Research and  Engineering (OUSD(R\&E)) under agreement number FA8750-15-2-0119. The U.S. Government is authorized to reproduce and distribute reprints for Governmental purposes notwithstanding any copyright notation thereon. The views and conclusions contained herein are those of the authors and should not be interpreted as necessarily representing the official policies or endorsements, either expressed or implied, of the U.S. National Science Foundation (NSF) or the U.S. Office of the Under Secretary of Defense for Research and Engineering (OUSD(R\&E)) or the U.S. Government. In addition, Y. Zhao, in the nursing program of Houston Community College, helped complete the error analysis of COVID-19 classification by using her medical imaging expertise.

\bibliographystyle{IEEEtran}
\bibliography{NIHGrant}
\end{document}